\documentclass[manuscript, acmlarge]{acmart}

\AtBeginDocument{%
  \providecommand\BibTeX{{%
    \normalfont B\kern-0.5em{\scshape i\kern-0.25em b}\kern-0.8em\TeX}}}

\setcopyright{acmcopyright}
\copyrightyear{2024}
\acmYear{2024}
\acmDOI{10.1234/1234567.1234567}

\acmJournal{IMWUT}
\acmBooktitle{}
\acmPrice{15.00}
\acmISBN{978-1-4503-XXXX-X/18/06}

\acmSubmissionID{7323}

\usepackage{threeparttable} 
\usepackage{caption}
\usepackage{subcaption}
\usepackage{siunitx}

\usepackage[shortlabels]{enumitem}

\usepackage{lipsum}
\usepackage[utf8]{inputenc}
\usepackage{graphicx}
\usepackage{wrapfig}
\usepackage{lscape}
\usepackage{rotating}
\usepackage{epstopdf}

\usepackage{geometry}
\usepackage{pdflscape}
\usepackage{afterpage}
\usepackage{lipsum}% dummy text
\usepackage{rotating}
\def\doitems{\def\item{\par
   \noindent\hbox to1.5em{\hss$\bullet$\hss}\hangindent=1.5em }}

\usepackage{threeparttable}
\usepackage{tabularx}
\usepackage{siunitx}

\usepackage{xcolor}
\usepackage{soul}

\usepackage{multirow}
\usepackage{fancyhdr}   %Kopfzeilen und co.
\fancypagestyle{firstpage}{     % define a custom header (Kopfzeile)
  \fancyhf{}
  \chead{\textcolor{gray}{This article has been conditionally accepted for publication in the proceedings of the \\ ACM on Interactive, Mobile, Wearable and Ubiquitous Technologies}}
  \fancyfoot[C]{\small{\textcolor{gray}{~\copyright~ © {Sizhen Bian, et al. | ACM} {2024}. This is the author's version of the work. It is posted here for your personal use. Not for redistribution. The definitive Version of Record is to be announced.}}}

}

\begin{document}

\title{Body-Area Capacitive or Electric Field Sensing for Human Activity Recognition and Human-Computer Interaction: A Comprehensive Survey}


\author{Sizhen Bian}
\email{sizhen.bian@pbl.ee.ethz.ch}
\affiliation{%
  \institution{ETH Zürich}
  %\streetaddress{P.O. Box 1212}
  \country{Switzerland}
}

\author{Mengxi Liu}
\email{mengxi.liu@dfki.de}
\affiliation{%
  \institution{DFKI}
  \country{Germany}
}

\author{Bo Zhou}
\email{bo.zhou@dfki.de}
\affiliation{%
  \institution{DFKI}
  \country{Germany}
}

\author{Paul Lukowicz}
\email{paul.lukowicz@dfki.de}
\affiliation{%
  \institution{DFKI}
  \country{Germany}
}

\author{Michele Magno}
\email{michel.magno@pbl.ee.ethz.ch}
\affiliation{%
  \institution{ETH Zürich}
  \country{Switzerland}
}

\renewcommand{\shortauthors}{Sizhen Bian, et al.}

\begin{abstract}

\thispagestyle{firstpage} 

Due to the fact that roughly sixty percent of the human body is essentially composed of water, the human body is inherently a conductive object, being able to, firstly, form an inherent electric field from the body to the surroundings and secondly, deform the distribution of an existing electric field near the body. Body-area capacitive sensing, also called body-area electric field sensing, is becoming a promising alternative for wearable devices to accomplish certain tasks in human activity recognition (HAR) and human-computer interaction (HCI). Over the last decade, researchers have explored plentiful novel sensing systems backed by the body-area electric field, like the ring-form smart devices for sign language recognition, the room-size capacitive grid for indoor positioning, etc. On the other hand, despite the pervasive exploration of the body-area electric field, a comprehensive survey does not exist for an enlightening guideline.  Moreover, the various hardware implementations, applied algorithms, and targeted applications result in a challenging task to achieve a systematic overview of the subject. This paper aims to fill in the gap by comprehensively summarizing the existing works on body-area capacitive sensing so that researchers can have a better view of the current exploration status. To this end, we first sorted the explorations into three domains according to the involved body forms: body-part electric field, whole-body electric field, and body-to-body electric field, and enumerated the state-of-art works in the domains with a detailed survey of the backed sensing tricks and targeted applications. We then summarized the three types of sensing frontends in circuit design, which is the most critical part in body-area capacitive sensing, and analyzed the data processing pipeline categorized into three kinds of approaches. The outcome will benefit researchers for further body-area electric field explorations. Finally, we described the challenges and outlooks of body-area electric sensing, followed by a conclusion, aiming to encourage researchers to further investigations considering the pervasive and promising usage scenarios backed by body-area capacitive sensing.

\end{abstract}

\begin{CCSXML}
<ccs2012>
   <concept>
       <concept_id>10003120.10003138.10003142</concept_id>
       <concept_desc>Human-centered computing~Ubiquitous and mobile computing design and evaluation methods</concept_desc>
       <concept_significance>500</concept_significance>
       </concept>
   <concept>
       <concept_id>10010147.10010257.10010293.10010294</concept_id>
       <concept_desc>Computing methodologies~Neural networks</concept_desc>
       <concept_significance>500</concept_significance>
       </concept>
   <concept>
       <concept_id>10010147.10010178.10010187</concept_id>
       <concept_desc>Computing methodologies~Knowledge representation and reasoning</concept_desc>
       <concept_significance>300</concept_significance>
       </concept>
 </ccs2012>
\end{CCSXML}

\ccsdesc[500]{Human-centered computing~Ubiquitous and mobile computing design and evaluation methods}
\ccsdesc[500]{Computing methodologies~Neural networks}
\ccsdesc[300]{Computing methodologies~Knowledge representation and reasoning}

% Keywords
\keywords{Electric field sensing, capacitive sensing, body-area network, body-area sensing, human activity recognition, human machine interaction, wearable. }

\maketitle

\section{Introduction}
\label{sec:introduction}

%\IEEEPARstart{S}%{ridging} the physical human behaviors and digital systems that assist individuals with targeted services like health monitoring and activity recording has been a ceaseless research topic for decades, especially since the presence of semiconductor technology~\cite{baltes1996future, ina1989recent} when silicon-based sensors were manufactured, benefitting from the integrated circuit design and manufacturing techniques, as presented by Dincer~et~al.~\cite{dincer2019disposable} with a historical timeline of key events in sensors. Naturally, humans experience the physical world, including their own behavior, through sensory organs like eyes and ears. To extend the perception capacities, a plentiful series of sensors are developed to perceive, monitor, or track objects that are likely imperceptible to the sensory organs. Human activity sensors help fill the gap between the physical world (human behavior and surroundings) and the electric systems that need an electric signal input and output specific feedback for individuals. 
Sensing human activities and presenting them in a digital format provide a better understanding of human behavior and the environment where that behavior happens, which benefits individuals on a large scale, including healthcare \cite{yue2020bodycompass, jones2021determinants}, social interaction \cite{di2018unobtrusive, gashi2019using}, life assistance \cite{zhang2023lt, kianpisheh2019face}, etc. A rich amount of sensing modalities have been explored to digitalize human activities for further inferences, as summarized in recent survey works \cite{fu2020sensing, bian2022state}. 
%The human activity sensing modalities can be sorted into five classes: field sensors, kinematic sensors, wave-based sensors, physiological sensors, and hybrid ones, according to the sensing principles behind them \cite{bian2022state}. % An in-depth description and comparison of the sensing tricks were presented. The targeted application was also summarized in three categories: body position-related, body action-related, and body status-related problems, aiming to supply researchers with a quick sensor decision in their specific use cases. 
%Besides those full-stack sensor surveys, 
Besides those comprehensive activity sensing surveys, researchers also focused on specific sensing modalities and supplied in-depth reviews, including their historical evolution, theories and models, critical techniques for applications, and data processing pipeline like various deep learning architectures, such as the works from Ma et al. \mbox{\cite{ma2016survey}}  on WIFI signal, Li et al. \mbox{\cite{li2019survey}}  on radar signal, Subhas et al. \mbox{\cite{mukhopadhyay2014wearable}}  on wearable sensors, Allah et al. \mbox{\cite{bux2017vision}}  on the vision-based sensor, and Wesllen at al. \mbox{\cite{sousa2019human}}  on the inertial sensor. Vision and inertial sensors are the most widely explored sensing modalities, both commercially and academically, in digitalizing body action knowledge. The former analyzes 2D or 3D images from optical sensors like RGBD and infrared cameras. Vision-based data are easy to collect/annotate compared to sensor-based data, which partly boosts the intensive exploration of vision-based solutions for human activity recognition (HAR) and human-computer interaction (HCI). The latter benefits from its properties of low power consumption, small form factor, and acceptable price. However, like all the other sensing modalities, despite their massive exploration, they are not winner-take-all and naturally face certain limitations, like occlusion and lighting conditions for vision-based and long-term drift for inertial sensor-based applications. In practical scenarios, various sensing solutions need to achieve mutual complementarity in joint development. Thus, researchers have explored a series of novel sensing techniques to address the shortcomings of available solutions and to extend the sensing ability of human behaviors and their ambient, like ultra-wide-band\cite{bouchard2020activity}, millimeter wave\cite{kwon2019hands}, and magnetic field\cite{bian2020social}. One of the massively explored, novel human action-related sensing modalities in the last decade is body-area electric field sensing, which provides non-intrusive, high-sensitive sensing ability for HAR and HCI. 

Table \mbox{\ref{Comparision_Field_Others}} roughly summarized the most commonly deployed sensing modalities in the field of human activity recognition and human-computer interaction, with performance metrics of cost, power consumption, sensing mode(active/passive), privacy issue, computing load, and robustness. The typical applications and a brief comment about their advantages and limitations are also supplied. Compared with other sensing modalities, body-area electric field sensing is the most cost-efficient one while supplying novel and extended sensing ability (a simple 555 timer or an instrumental amplifier is enough for setting up a sensing front end with high sensitivity to body motion or environment variation). Although it faces the robustness issue, especially in the passive sensing mode, it enjoys the advantages of low power consumption and low computing load, as a result of its simple hardware implementation and low dimensional signal.

%With the overwhelming development of neural network-based algorithms for image analysis, the vision-based solutions outperform with their performance in accuracy and generality. The latter benefits from its properties of low power consumption, small size, and acceptable price. Most consumer electric devices, like smartphones, smartwatches, and \Gls{AR} and \Gls{VR} equipment, embed the \Gls{IMU} as a critical(or even the unique) sensor for motion sensing. However, like all the other sensing modalities, despite their massive exploration, they are not winner-take-all and naturally face certain limitations, like occlusion and lighting conditions for vision-based and long-term drift for inertial sensor-based activity recognition tasks. In practical applications, various sensing solutions need to achieve mutual complementarity in joint development. Thus, researchers have explored a series of novel sensing techniques to address the shortcomings of available solutions and to extend the sensing ability of human behaviors and their ambient over the past decades, like ultra-wide-band\cite{bouchard2020activity}, millimeter wave\cite{kwon2019hands}, and magnetic field\cite{bian2020social}.

%One of the massively explored, novel human activity-related sensing modalities in the last decade is capacitive sensing in the body area, providing non-intrusive, motion-sensitive sensing ability for HAR and HCI. 

\subsection{Background}

Body-area electric field sensing, also defined as body-area capacitive sensing, makes use of the conductivity character of the human body and integrates the whole body or body part into an electrical system for context perceiving in the field of human-computer interaction and human activity recognition. Over the last decade, researchers have explored plentiful systems for wearables and smart environments backed by body-area capacitive sensing, like the smart glasses equipped with capacitive sensors for facial expressions and head gesture recognition \mbox{\cite{matthies2021capglasses}}, which makes use of the proximity of skins to the electrodes that are sensitive to surrounding motions; The wristband designed for exercise recognition \mbox{\cite{bian2019passive}} simply by passively sensing the electric field variation 

\afterpage{%
\newgeometry{margin=2.0cm} % modify this if you need even more space
\begin{landscape}
\begin{table}[]

%\captionsetup{justification=centering,labelsep=newline,textfont={footnotesize,sc},labelfont=footnotesize}
\centering
%\begin{threeparttable}
\caption{ Sensing Modalities in HAR and HCI Tasks}
\label{Comparision_Field_Others}

\begin{tabular}{ p{2.0cm} p{1.5cm} p{1.5cm} p{1.5cm}  p{1.0 cm} p{1.0cm} p{1.5cm}  p{5.0cm} p{3.5cm}  p{1.5cm} }

\toprule
Modality  & Cost (USD) & Power Level & Active/ Passive & Privacy Issue & Compute Load & Robustness & Typical Application & Comment & Example Works\\ 
\midrule
WiFi  & tens  & $\approx$tens Watt & active   & no & medium & low & positioning, ADL(activity of daily life), ambient intelligence  & pervasiveness,  environmental sensitivity & \cite{wang2015understanding, zhang2022wi} \\\midrule
UWB  & tens  & $\approx$mW & active    & no & low & low  & positioning, proximity, ADL, gesture recognition, ambient intelligence & multi-path resistive, high accuracy, costly for massive consumer usage & \cite{du2020segmented, khan2018human} \\\midrule

mmWave  & tens  & $\approx$W & active    & no & medium & low  & positioning, proximity, ADL, gesture recognition, health monitoring, ambient intelligence  & high accuracy, low power efficiency for massive consumer usage & \cite{patra2018mm, ru2023sensing} \\\midrule

Ultrasonic  & hundreds  & $\approx$mW to
 W & active    & no & low & low &  positioning, proximity, ambient intelligence  & high accuracy, weak robustness & \cite{ghosh2017automatizing, fan2022development}  \\\midrule

Optic  & tens of hundreds   & $\approx$W and above & passive   & yes & high & medium &  positioning, proximity, ADL, gait analysis, gesture recognition, surveillance & comprehensive approach, high resource consumption & \cite{park2016depth, ahmed2021hand} 
 \\\midrule

ExG  & hundreds  & $\approx$tens mW & passive  & no & medium & high  & sports, healthcare monitoring,  ADL &  high resolution, noise sensitive & \cite{nurhanim2021emg, qi2019intelligent}  \\\midrule

IMU  & a few & $\approx$mW & passive & no & high & medium & positioning, ADL, gesture recognition, healthcare monitoring, gait analysis,  sports & dominant sensing modality, accumulated bias & \cite{zhuang2019design, kim2019imu} \\\midrule

Magnetic Field(DC) & tens & $\approx$mW & passive  & no & low & high & proximity, gesture recognition & high accuracy, short detection range & \cite{maekawa2013activity, yi2022magnetic}\\ \midrule

Magnetic Field(AC) & tens & $\approx$hundreds mW & active  & no & low & high &  positioning, proximity, & high robustness, limited detection range & \cite{bian2020social, bian20233d}  \\ \midrule

Electric Field(active) & a few & $\approx$mW & active    &no  & low & low &  positioning, proximity, ambient intelligence & high sensitivity, noise sensitive & \cite{bian2021capacitive, zhou2023mocapose} 
 \\ \midrule

Electric Field(Passive) & a few & $\approx$sub-mW & passive   &no  & low & low &  positioning, proximity, sports, gait analysis,  ambient intelligence & high sensitivity, noise sensitive & \cite{cohn2012ultra, tang2019indoor} \\ 

\bottomrule
\end{tabular}
\end{table}
\end{landscape}
\restoregeometry

}

\restoregeometry

\noindent
between the human body and environment during different body motion patterns. On one hand, the human body itself carries charges, thus inherently forming an electric field to the surroundings. By monitoring the field variation in either direct or non-direct approaches, the body motion or surrounding change could be deduced. On the other hand, the intrusion of the human body into an existing electric field will result in deformation of the existing electric field; such a deformation could be perceived for intrusion pattern recognition. In nature, charges carried on living bodies play an important role in survival. For example, some special fishes like the gymnarchus niloticus \mbox{\cite{lissmann1958mechanism, bullock1982electroreception}} utilize the charge variation on the body surface to sense the surrounding risk: the potential distribution over the surface of the fish is detected by a series of receptors; this information is then interpreted to indicate the position of objects with a conductivity differing from that of water.

Capacitive sensor was born in the pioneering days of electricity and was used to store the charge in the form of Leyden Jar \mbox{\cite{heilbron1966gm}}, as electricity was thought to be fluid in the 18th century. In the late 19th century, paper capacitors and electrolytic capacitors were invented and were used from the early 20th century. During the second and third industrial revolutionaries, different forms of capacitors were invented to satisfy the rapid development of industry. With the development of capacitive sensors, a rich amount of capacitive-based applications were explored. Meanwhile, the body-area capacitive sensing also started to be explored, like the capacitive blood pressure manometer \mbox{\cite{tompkins1949new}}, which senses the capacitance variation caused by the varied blood pressure in the arteries. The first body capacitance-enabled consumer device is the special music instrument Theremin \mbox{\cite{nikitin2012leon}}, invented by the 23-year-old Leon Theremin in 1919 by accident. In 1965, E.A. Johnson invented what is generally considered the first finger-driven capacitive touchscreen \mbox{\cite{johnson1965touch}}, utilizing the conductance/capacitance property of the finger. Starting from the 21st century, body-area capacitive sensing showed an explosive development benefitting from its pervasiveness and the emerging novel and high-precision sensing approaches, such as the resonator-based capacitance chips like FDC2x1x from Texas Instruments(TI), and the charger variation-based capacitance chips like QVAR from STMicroelectronics. Since capacitive sensing is a low-power, low-cost, contactless sensing technique, it was applied to applications like biophysical signal monitoring \mbox{\cite{yama2007development}}, position tracking \mbox{\cite{osoinach2007proximity}}, activity classification \mbox{\cite{cheng2010active}}, and intrabody communication \mbox{\cite{shinagawa2004near}}, etc. In Appendix \mbox{\ref{History_appendix}}, we comprehensively described the historical information about the development of capacitance, including the different capacitor forms, capacitive-based sensors, and their backed applications, with a focus on body-area applications (in bold format), as summarized in Fig. \mbox{\ref{History}}.

The historical diagram and the following survey are obtained by thoroughly exploring the search engines like IEEE Xplore, Google Scholar, ACM digital library, etc. We searched through peer-reviewed conference and journal papers, books, dissertations, and patents. Since there are multiple terms for capacitor and body-area capacitance, we use the following keywords (and their combination) to compile the databases through an iterative process of research: "capacitor/capacitive", "condenser" (the term ‘capacitor’ was referred to as "condenser" in history and didn’t start being used until sometime in the 1920s), "sensor/sensing", "transducer", "(static) electric field", "body capacitance", etc. For this aspect, this research also aims to provide a common terminology that will serve as a unifying framework for all engineering disciplines involved in body-area capacitive/electric field sensing.

\subsection{Related surveys}

Despite the ubiquitous role of body-area capacitive sensing, especially on wearables and instrumented environments for both HAR and HCI, there are few related surveys on this topic. By thoroughly exploring the literature, we found only one review and one survey related to body-area capacitive sensing, and both were published in 2017. 
%With the title of "Finding common ground: A survey of capacitive sensing in human-computer interaction",
Tobias et al. \cite{grosse2017finding} strove to unify capacitive sensing by advocating consistent terminology and proposing a new taxonomy to classify capacitive sensing approaches, considering that the broad field of capacitive sensing research has become fragmented by different approaches and terminology used across the various domains. They also provided an analysis and review of past research and identified challenges for future work, aiming to create a common understanding within the field of HCI. The second is a review paper from Arshad et al. \cite{arshad2017review}, %titled "Electric Field Sensing for Human-Computer Interaction Applications". The paper 
which provided the principle of the electric field sensing for HCI, categorized the sensing mode, surveyed the related works, and gave the reason why the growth in this field is desirable or even necessary. 

However, both surveys focused only on the capacitance/electric field around the whole body and applications in the domain of HCI. The electric field emitted from different body parts and body-to-body electric field are not described. Besides that, a thorough review of related hardware implementation, data processing, and the challenge that limits the pervasive deployment are still missing.

\begin{figure}[h]
\centerline{\includegraphics[width=0.55\textwidth, height=4.0cm]{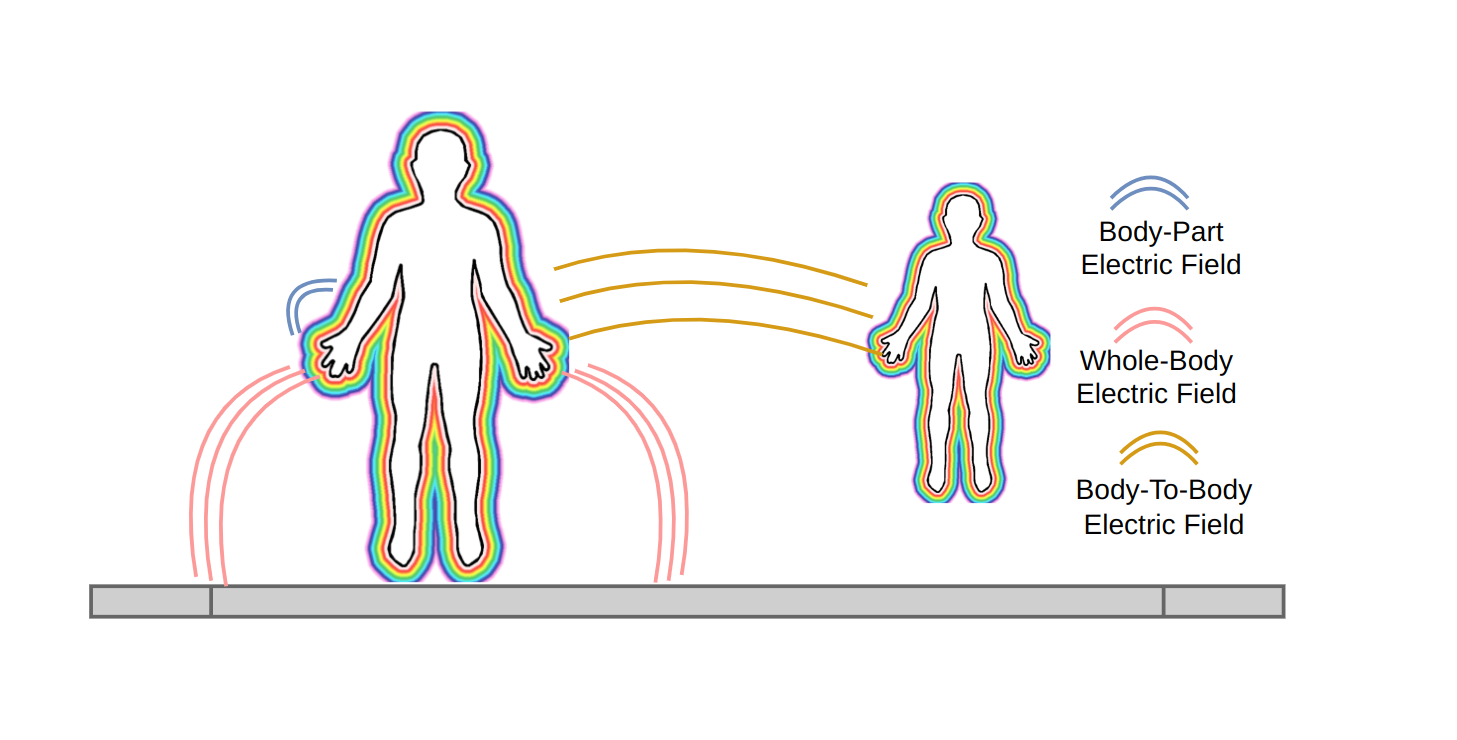}}
\caption{Three groups of work on the body-area electric field: sensing the body-part electric field (e.g. for gesture recognition) which mostly leverages the deformation of the existing instrumented electric field caused by the intrusion of that body-part; sensing the whole-body (or body-to-ground/environment) electric field (e.g. for motion detection) where a passive electric field varies during body-ground/environment action; sensing the body-to-body electric field for group interaction and cooperation recognition}
\label{Three_group}
\end{figure}

\subsection{Paper Aims and Contribution}
Due to the feature that the body can, first, radiate a passive electric field from the body to the surroundings, second, deform an existing electric field in instrumented surroundings or wearables, the exploration of body-area electric field or capacitive sensing has shown an increasing trend for researchers who majored in the field of HAR and HCI, and many impressive works have been published in the past two decades, ranging from single body-part capacitive-based motion sensing \cite{matthies2017earfieldsensing, luo2020eslucent, braun2014towards} to the whole body motion tracking \cite{zhou2023mocapose} and body-to-body capacitive-based cooperation sensing \cite{canat2016sensation, bian2022contribution, staudt_pascal_2022_6798242}. Some designs reported impressive activity recognition accuracy, like in body posture detection \cite{braun2015capseat}, and ultra-low power consumption for wearable motion sensing \cite{reinschmidt2022realtime}, which is over ten times power saving than the traditional IMU-based motion sensing. However, despite the pervasive exploration of the body-area electric field, a thorough survey is still missing to comprehensively describe the development of body-area capacitive sensing and supply an enlightening guideline for this technology. This survey tries to fill the gap by supplying a full-stack overview of body-area capacitive sensing in HAR and HCI. For this purpose, we first categorized the past research works on body-area electric field sensing into three classes depending on the involved body form, aiming to provide a clear application map for readers. An in-depth analysis of the backed sensing principle, hardware implementation, applied algorithm, targeted application, performance, and limitation, of each typical work within each category is supplied. We then summarized the sensing source signals, the corresponding analog front ends, and the data processing pipeline, aiming to provide essential information so that other researchers can decide if and in what capacitive sensing form and data processing approach are suitable for their specific applications. Finally, we described the challenges for the massive deployment of the body-area capacitive sensing technique in practical applications like the weakness in robustness, and proposed several directions for a more accurate, robust, easy-of-use sensing with body-area capacitance, aiming to encourage researchers for further novel investigations considering the pervasive and promising usage scenarios backed by body-area capacitive sensing. Body-area capacitive sensing offers a simple and natural sensing approach while being very cheap and efficient, this survey will contribute to researchers from both HAR and HCI communities and stimulate and facilitate future research on this topic.

This paper is structured as follows: in Section \mbox{\ref{sec:Taxonomy}}, we summarized a few terminologies that commonly appear in related papers. Section \mbox{\ref{sec:BodyForm_survey}} gave an in-depth analysis of recent works based on the body-area electric field. For each work, we summarized the sensed object, hardware form, sensing front-end, subject of the work, applied algorithm and its performance, contribution, and limitation. Section \mbox{\ref{sec:pipeline}} focused on the three kinds of hardware implementation and the data processing pipelines, aiming to ease other researchers with the same research target by enlightening them with novel sensing and data processing ideas.  In Section \mbox{\ref{sec:challenges}} and \mbox{\ref{sec:Outlooks}}, we described the challenges and outlooks of body-area electric field sensing. Here we focused on the limitations of robustness and generalization of the subject towards a ubiquitous deployment. In the outlook section, we proposed a few future development directions, like active shielding in hardware and continuous learning in algorithms to address the limitations  Section \mbox{\ref{sec:Conclusion}} concluded our survey work. The main contributions of this survey work can be summarized as follow:

\begin{enumerate}
    \item We summarized the development of capacitance and related sensors and applications in history and described a few taxonomies to classify  body-area  capacitive sensing approaches, which also helps readers with a better understanding of sensing principles. 
    \item We categorized body-area capacitive sensing into three classes: body-part capacitive sensing, whole-body capacitive sensing, and body-to-body capacitive sensing, as Fig. \ref{Three_group} depicts, aiming to supply the readers with an in-depth map of body-area capacitive applications. We enumerated broadly the published works within each category. An in-depth description of the underlying technical tricks is given. For typical works, we summarized the sensing hardware form, sensing/operation mode, source signal, sensing front end, applied algorithm, target performance, contribution, limitation, etc., in an attempt to explain the current state of understanding on the topic. 
    \item We summarized the circuits/components developed for body-area capacitive sensing and the data mining method for HAR and HCI so that readers could be lightened with further implementing approaches or choose the best one for their specific applications. 
    \item We analyzed the challenges for the massive deployment of the capacitive sensing technique, such as the lack of robustness, and proposed potential future directions of the body-area capacitive sensing based on our years of experience in this domain, hoping to trigger more novel ideas for promoting the development of body-area capacitive sensing. 
\end{enumerate}

\section{Taxonomy}
\label{sec:Taxonomy}

Body-area capacitive sensing is a kind of field sensing technique with the potential of being a low-cost, power-efficient, and non-intrusive HAR and HCI solution that is ideal for long-term and wearable sensing scenarios. This section describes a few terminologies that commonly appear in related papers, describing the principles of this specific field sensing in different aspects.

\subsection{Active and Passive sensing mode}

%\begin{figure}[t]
%\centerline{\includegraphics[width=0.5\columnwidth]{Figures/PassiveActive.png}}
%\caption{Active and passive body-area electric field sensing}
%\label{PassiveActive}
%\end{figure}

\begin{figure}
\begin{minipage}[t]{0.45\linewidth}
\centering
%\raggedleft
\includegraphics[width=0.9\textwidth,height=4.0cm]{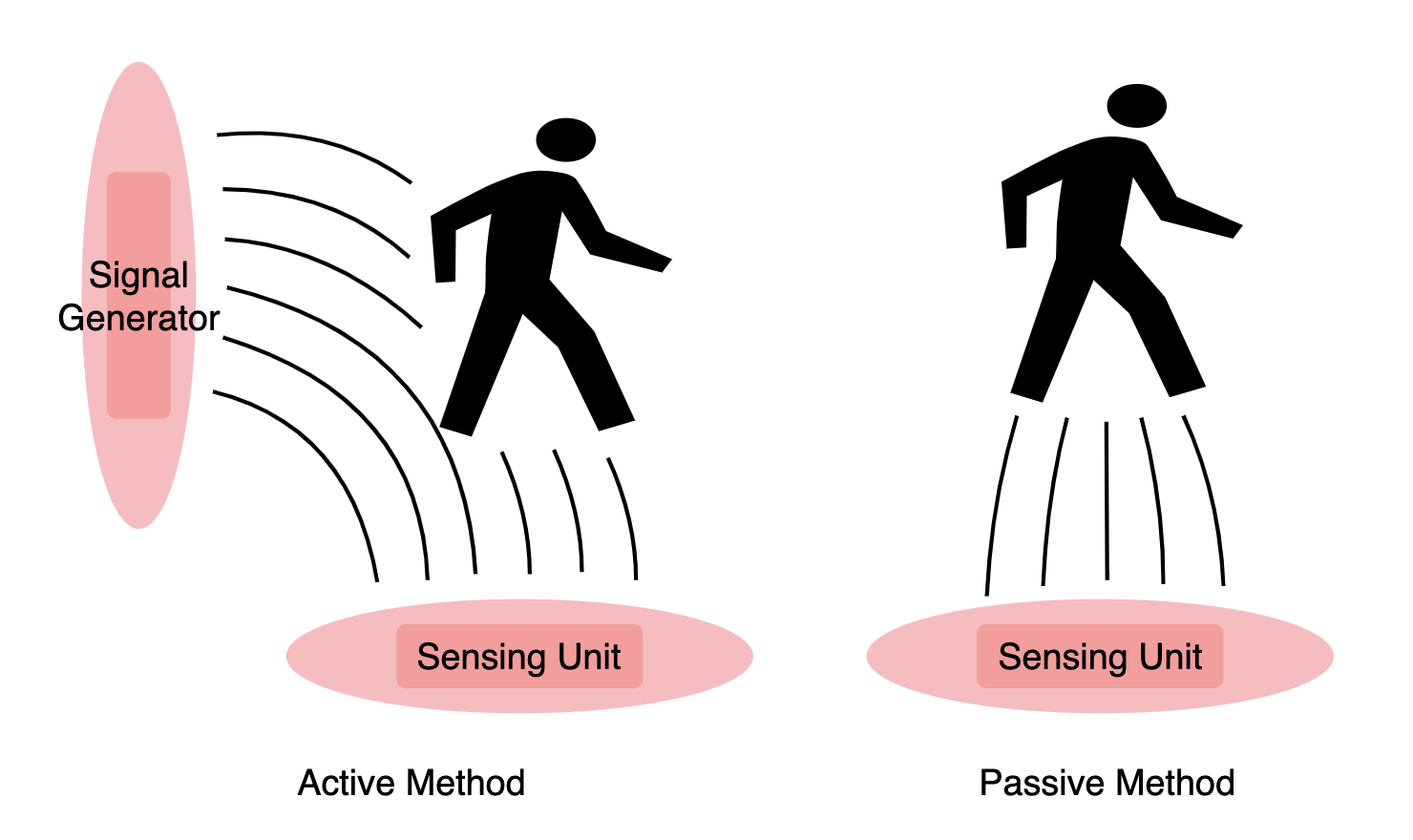}
\caption{Active and passive body-area electric field sensing}
\label{PassiveActive}
\end{minipage}
%\quad
\quad
\begin{minipage}[t]{0.45\linewidth}
\centering
%\raggedright
\includegraphics[width=0.7\textwidth,height=5.0cm]{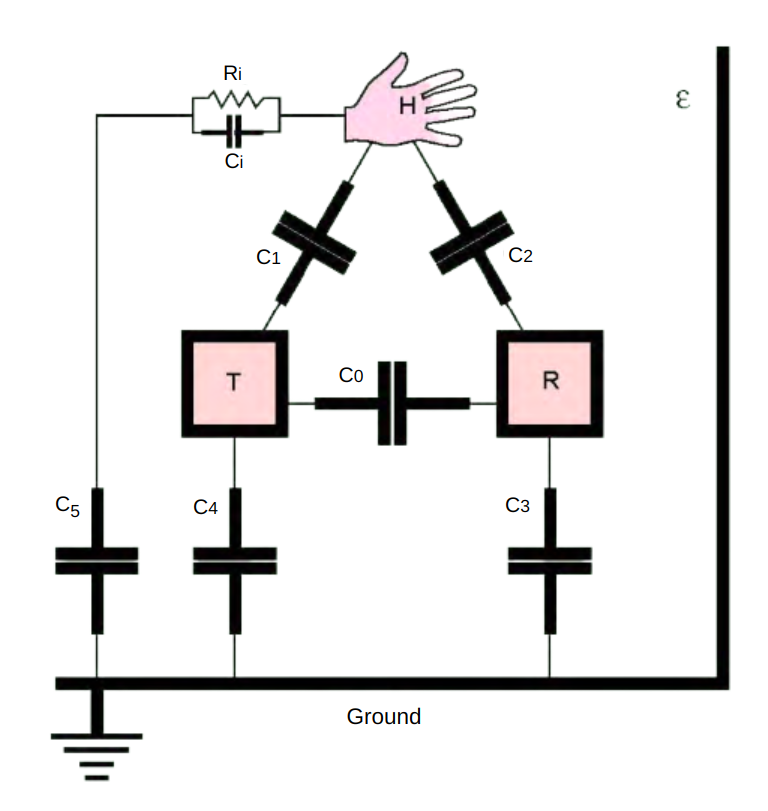}
\caption{Lumped circuit model of body-area electric field sensing}
\cite{smith1996field}
\label{CircuitModle}
\end{minipage}
\end{figure}

Active and passive sensing differs from the signal source. An active sensor is a sensing device that requires an external power source to generate the signal. In contrast, a passive sensor detects and responds to a signal that is derived from the natural environment. Figure \ref{PassiveActive} shows the body-area electric field sensing in both modes. In an active method, a pair of AC/DC electric field generator/transmitter and receiver exists, and the field distributes in between. The human body acts as an interference source distorting the field strength, resulting in signal variation at the sensing unit. In the passive mode, the sensing unit senses the field that exists naturally, like the static electric field from the whole body and the working appliance in a living environment. Since the passive mode doesn't rely on a specific electric field, it is more susceptible and less robust than the active mode \cite{matthies2021capglasses}, but enjoys the advantage of ubiquity and less power consumption regarding the sensing hardware.

\subsection{Load, Transmit and Shunt operating mode}

%\begin{figure}[!t]
%\centerline{\includegraphics[width=0.5\columnwidth]{Figures/SensingMode.png}}
%\caption{Lumped circuit model of body-area electric field sensing
%\cite{smith1996field}}
%\label{CircuitModle}
%\end{figure}

Applying capacitive sensing to the body action- and status-related applications (touch, positioning, posture) was first introduced by Zimmerman et al. in 1995 \cite{zimmerman1995applying}, where two sensing modes were proposed: shunt mode and transmit mode. In 1996, J. R. Smith \cite{smith1996field} from the same team presented the third mode: loading mode, as Figure \ref{CircuitModle} depicts with a lumped circuit model, where a pair of transmitter(T) and receiver(R) was used to abstract the capacitances around (electric field supplier and sensor), and a hand(H) was used to symbolize the object that interacts with the transceiver pair. $R_i$ and $C_i$ represent the internal electric properties of the body. $C_5$ is the capacitance between the body and the environment, mainly the ground, which is experimentally observed \cite{bian2021systematic} to be affected by a series of factors like sole materials/heights, postures, surroundings, etc. $C_0$ to $C_4$ are capacitances among the involved items, formed as a result of potential differences. The distance variation between the object and transceiver will cause capacitance variation, which is then being measured directly or indirectly to interpret the body activities and interactions. 

In load mode, the transmitter alone plays both the role of electric field supplier and the role of field sensing unit. Thus no receiver exists in this mode. When the object approaches the transmitter, $C_1$ will increase, and this variation will cause charge redistribution or resonant frequency drift at the transmitter unit. Theremin, for example, senses the proximity of hands and utilizes the resonating frequency drift for the hand-antenna distance indicator. The varied frequency is then encoded to generate the sound. 

In transmit mode, the object is located close to the transmitter, either strongly coupled or directly contacted, meaning that $C_1$ is much larger than $C_0$ (and sometimes also  $C_2$), but will not influence the sensing sensitivity of $C_0$ when the object approaches the receiver or make actions near the receiver. As Rekimoto et al. \cite{rekimoto2001gesturewrist} presented the GestureWrist, with all electrodes around the wrist, the wrist shape change during a gesture action will cause the capacitance variation ($C_0$) between the transmitter and receiver while keeping the $C_1$ and $C_2$ relatively stable.   

In shunt mode, the object has no strong connection to the transmitter. Without the object being near or in between, the electric field between the transmitter and receiver ($C_0$) is stable. When the object intrudes the electric field, $C_1$ and $C_2$ will increase, which can be equivalent to the increase of $C_0$, since the object could be regarded as a conductive medium between the transmitter and receiver. An example is the Wall++ designed by Zhang et al. \cite{zhang2018wall++}, allowing walls to become a smart infrastructure by instrumenting the walls with diamond electrode patterns. Thus room-scale interactive and context-aware applications, like the recognition and tracking of touch and posture, were enabled. 

Such taxonomies about operating mode have been widely adopted and explained by researchers in their works \mbox{\cite{fu2020sensing, arshad2017review,grosse2013opencapsense, matthies2021capglasses, zeeman2013capacitive}}, including the survey paper from Tobias et al.\mbox{\cite{grosse2017finding}}, where a fourth operating mode is proposed with the name of receive mode when the human body is very closely coupled to the receive electrode, which was normally counted in shunt mode in other works \mbox{\cite{zimmerman1995applying, smith1996field, grosse2012enhancing, braun2015capacitive, fu2020sensing}}. To provide a set of the most commonly named operating modes, we count the receive mode into the shunt mode in this survey.

\subsection{Frequency, Current and Time sensing sources}

\begin{figure}[t]
\centerline{\includegraphics[width=0.8\columnwidth]{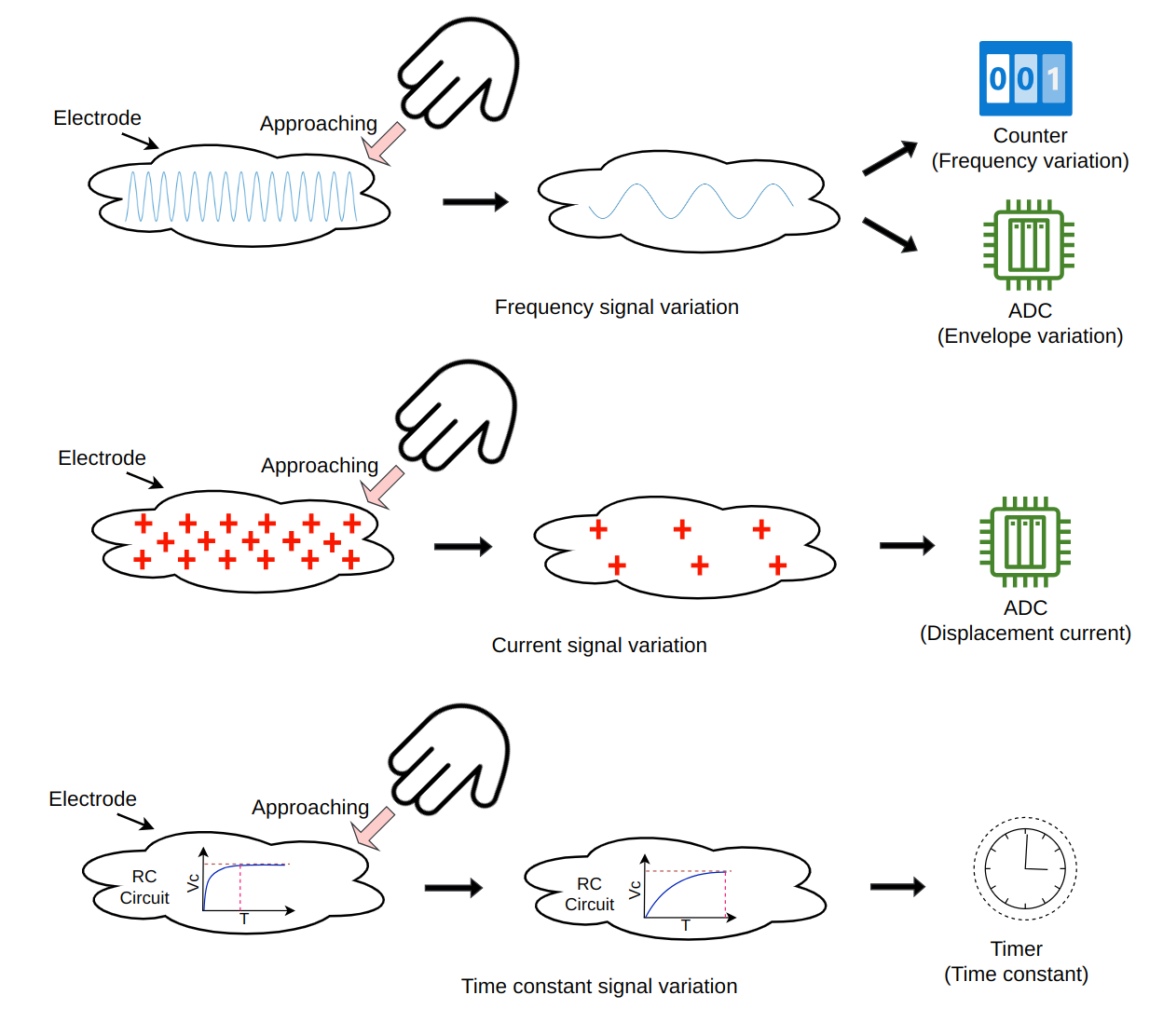}}
\caption{Sensing sources of body-area electric field sensing: frequency, Current and Time}
\label{SensingPrinciple}
\end{figure}

Since capacitance itself can not be sensed directly, to acquire the body-area capacitance signal, some other physical quantities, which will respond to a capacitance variation and can be directly recorded by digital devices, are used to indicate the capacitance variation. Depending on the electric field type, researchers designed different sensing circuits and explored mainly three physical parameters being measured for further digital processing. 

The frequency signal is the most commonly used signal source, where an AC signal is triggered on the electrode (usually by a resonant tank), generating an alternating electric field. The frequency or envelope of the sensed alternating signal on the sensing unit varies when the object approaches the electrodes and is acquired by the controller using a counter or AD module. For example, \cite{cheng2010active} used the Colpitts oscillator as the sensing front end, and sensed the envelope of the oscillating signal after an amplification step with an AD module. Since frequency signal is less susceptible to environmental static electric field noises, this sensing approach is more robust than the others. 

The second commonly used signal source is the current signal, namely the charge flow, through voltage measurement. When the electrode has a potential difference to the environment, a static electric field is formed. An approaching conductive object, like the hand of a human body, will result in the charge re-distribution. The flow of charge can be represented as the potential variation in forms of charging or discharging. Thus by observing the potential variation by simply an ADC module, the capacitance of the electrode to the surrounding can be revealed. \cite{bian2019passive}, for example, designed a front end with simply a few discrete components followed by a high-resolution ADC module. The front end supplies charges to the electrode, which is in direct contact or strong coupling with the body. When the body makes movements or a second body approaches, the capacitance of the electrode to the surrounding varies, which results in the potential variation of the electrode. Since the current signal source doesn't need a signal-triggering circuit, the current-based sensing unit enjoys the advantage of ultra-low power consumption, as claimed in \cite{cohn2012ultra}. It is also the most sensitive approach for capacitance variation observation, a minor charge flow will be mapped into a potential shift when a high input impedance exists in the sensing front end.

The third signal source that can be used for capacitance variation representation is the time constant in an RC circuit. The time constant indicates the charging or discharging time of a capacitor in an RC circuit. An approached hand will enlarge the capacitance from the electrode to the surrounding. Thus, a larger time constant value can be measured by a controller's timer. Although this approach also observes the charging and discharging process on the electrode, it differs from the current signal approach since this approach needs to actively activate the charging and discharging process, by simply pulling up the voltage on the electrode or pulling it down to the ground, so that a sequential of time constant value can be obtained. In \cite{matthies2021prototyping}, the authors built a capacitive prototype for facial and head gesture detection by measuring the time constant value of the RC circuit. Compared to the other signal sources, this method is easy to accomplish. A general microcontroller and a single resistor that connects two general-purpose input and output pins are sufficient for a sensing front end.

\section{Electric Field Sensing with different involved body form}
\label{sec:BodyForm_survey}

This section will thoroughly survey the research works based on the body-area electric field. To comprehensively present the current exploration status of the body-area electric field in HAR and HCI and to give readers a full-stack understanding of related works, the following elements are picked from each work for summarisation: 
\begin{itemize}
    \item The sensed object: hand, wrist, body, etc.
    \item The hardware form: glass, seat, ring, etc.
    \item The subject of the work: gesture recognition, gait analysis, etc. 
    \item Sensing mode: passive, active. 
    \item Operation mode: shunt, transmit, load.
    \item Sensing source: frequency, current, time.
    \item Core sensing front end: different kinds of capacitive chips or customized design. 
    \item Applied algorithm for the subject of the work: SVM, CNN, LSTM, etc. 
    \item The evaluation and performance.
    \item The contribution and limitation.
\end{itemize}

Since body-area electric field sensing applications cover a wide range of topics,
it is necessary to sort the related topics impressively and compactly.
Depending on the sensed object and the present area of the utilized electric field (either naturally existing or intentionally generated), which is mainly decided by the working principle and the deployment of the sensing units, we sorted three groups of body-area electric field sensing: the body part electric field, the whole body electric field, and the body-to-body electric field. For the body part electric field works, we sorted them regarding different organs that are being sensed, as this will supply readers with a better overview of the various (either naturally existing or manually generated) electric fields across the body. Meanwhile, such sorting will ease our documentation work. For the whole body and body-to-body electric field works, as the utilized electric field is fixed (the electric field between body and ground or between body and body), we sorted the investigated works according to the different applications.

\begin{figure*}[h]
\centerline{\includegraphics[width=0.6\textwidth, height=7.0cm]{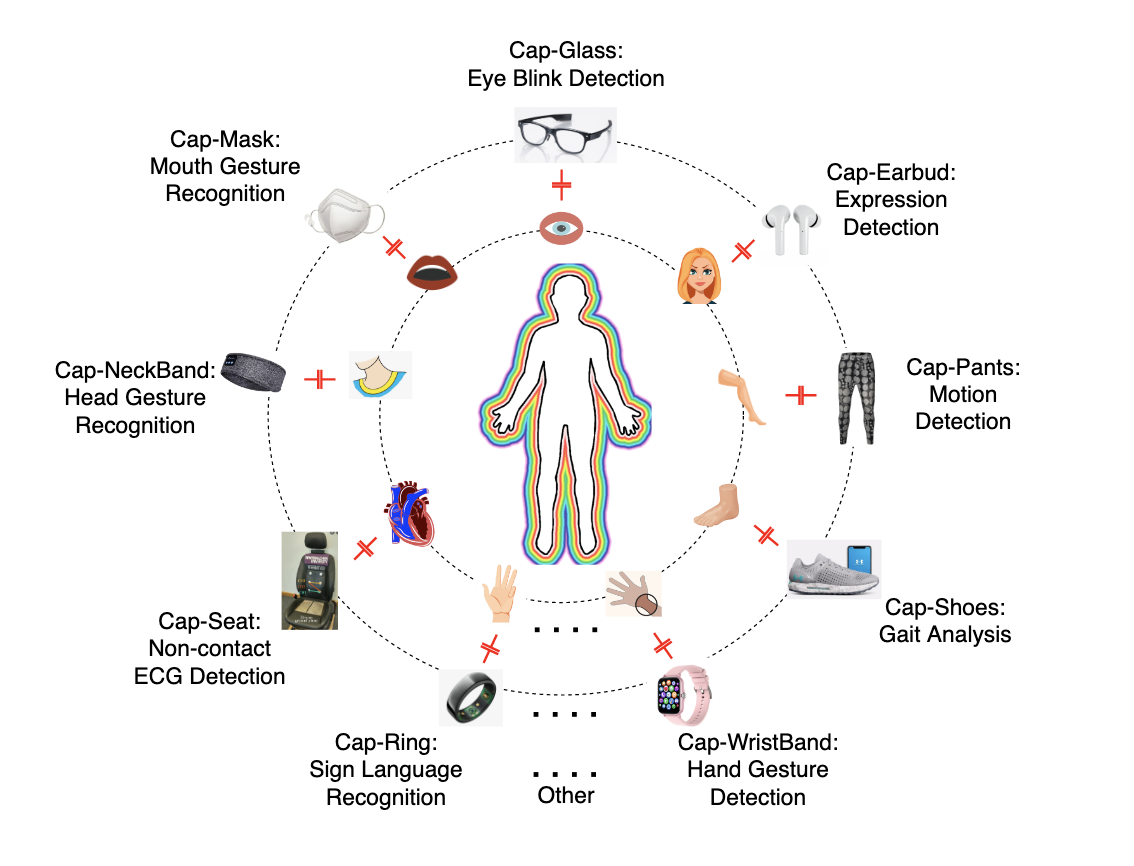}}
\caption{Applications leveraging the electric field existing/generated near the various body tissues}
\label{Body_Part}
\end{figure*}

\subsection{Body-part electric field sensing}

Body part electric field sensing covers the topics that aim at an organ of the body as the sensing object, seeking to obtain the body part's internal electrical property or action pattern with either active or passive electric field sensing mode. This group is the most widely explored one because of the pervasive usage scenarios in wearables. Figure \ref{Body_Part} selected the topics that leverage the active or passive electric field near different body parts with different carriers for a specific task. Displayed are the commonly explored ones in the literature, like capacitive hand gesture recognition, capacitive gait analysis, etc. Table \ref{BodyPartTable} and Table \ref{BodyPartTable2} summarized the works that have been published in recent ten years utilizing the body-part electric field sensing. The following body parts were explored with electric field sensing modality in those works:

\subsubsection{Face}

Facial expression recognition has been a hot topic in recent years, triggered by the development of virtual reality. Most existing solutions are based on deep visual processing \cite{li2020deep}. To overcome the limitation of the visual approach, like the light condition and computational load, researchers explored sensor-based solutions like stretchable sensor \cite{masum2021stretch}, acoustic sensing \cite{song2022facelistener}, etc. Denys et al. \cite{matthies2021prototyping} developed a capacitive-based smart glass with multiple electrodes to sense facial expressions. Since the muscle movement under different expressions will cause proximity variations from the electrode to facial muscles, a varied array of capacitance values could thus be used for facial expression detection. with a random forest classifier, the authors claimed accuracy of 91.25\% with 12 facial and head gestures recognition. Although the described performance was only achieved in labor, the work showed the feasibility of capacitive sensing for facial expression recognition with low-cost, low power consumption, and low computational load. 

\subsubsection{Eyes}

Eye blink rate has been demonstrated to be an efficient indicator of fatigue, mental load, and other cognitive capacities. The most common electrooculography(EOG)-\cite{reddy2011analysis} and electroencephalogram(EEG)-\cite{agarwal2019blink} based eye blink rate detection approach requires high resolution, specialized physiological signal detection chips to measure the subtle potential caused by eye movement. In \cite{luo2020eslucent}, the authors utilized capacitive sensing for blink detection, where a customized eyelid sticker was designed with a multi-layer structure, enabling a capacitance variation detection functionality when the eyelid moves. An average detection precision of 82.6\% was presented with intentional and voluntary blinks. To address the limitation of on-skin deployment, Liu et al. \cite{liu2022non} presented a non-contact glass-based eye blink detection prototype, where the active capacitive chip FDC2214 was used to detect the eyelid action caused capacitance variation near the single-end electrode. The principle behind it is much like the Theremin, where the frequency signal from a resonant circuit is utilized to observe the proximity of a conductive object. Depending on the resolution of the oscillator-based capacitive sensor, very subtle intrusion in the electric field of the electrode will cause a noticeable resonant frequency shift. Since the frequency shift data is just straightforward, one-dimensional continuous data, an adaptive threshold could be used to detect the eye blink in real time easily. The authors reported an average precision of 92\% and recall of 94\% with data from eight volunteers and five scenarios, demonstrating the feasibility and robustness of the proposed solution. However, the authors also stated that to keep a reliable detection performance, firstly, the glass frame needs to match the head form, aiming to keep a close distance of the eyelid and electrode on the frame; secondly, intensive body actions like running should be avoided, since the minor vibration of glass frame will also cause impressive capacitance variation of the electrodes.

\subsubsection{Ears}

Consumers growing emphasis on health and fitness is boosting the demand for smart devices. Compared with smartwatches and fitness bands, the interest in smart earbuds has shown a faster growth rate in recent years, especially the true wireless stereo(TWS), benefiting from the ultra-low power Bluetooth technology. As a result, more sensors were embedded into the earbud for high-level context analysis, like cough detection with IMU\cite{zhang2022coughtrigger} and heartbeat detection with optical sensor \cite{boukhayma2021earbud}. In \cite{matthies2017earfieldsensing}, Denys et al. explored the possibility of facial expression recognition with capacitive sensing in the form of an earbud. They designed a customized passive electric field sensing front end, composed of a few discrete components, two amplifiers, and one instrumental amplifier (the adoption of it depends on the sensing mode: single mode or differential mode). The electrodes in the front end sense the electric field change caused by facial muscle movements. Three other off-the-shelf capacitive sensing front ends were also used for comparison. Two studies were performed to reveal how electric
sensing technologies could perform when using an electrode in-ear plug. Finally, with five facial gestures, the customized front end achieved a precision of 90\% while sitting and 85.2˜\% while walking with a J48 decision tree classifier. Although the volunteer's number was limited, the work presented a novel facial gesture recognition solution by sensing the electric field inside the ear canal.

\subsubsection{Legs}

Legs are the most critical body part in locomotion and are ideal for proximity observation with capacitive sensing since leg-dominated actions characters larger motion scales than the other body part-dominated actions.
In \cite{haescher2015capwalk}, the authors deployed a capacitive sensing band on the leg to recognize five walking styles:

\afterpage{%
%\noindent
\newgeometry{margin=2cm, left=3.5cm} % modify this if you need even more space
%\newgeometry{left=1cm}

\begin{landscape}
\begin{table}[htbp]
\caption{Body-part electric field applications(1)}
\label{BodyPartTable}
%\renewcommand{\arraystretch}{1.4}
%\begin{tabularx}{\textwidth}{@{}l*{8}{C}c@{}}
\footnotesize
\begin{tabular}{ p{1.2cm} p{0.5cm} p{0.6cm} p{1.7cm}  p{0.8 cm} p{0.8cm} p{0.9cm} p{1.2cm} p{1.0cm} p{3.8cm} p{8.0cm}}
\toprule
References-Year   & Body Part & Form & Subject & Passive/ Active & Sensing Mode & Source Signal & Hardware & Algorithm & Performance & Contribution and Limitation  \\ 
\midrule
\cite{matthies2021prototyping}-2021   & Face      &  Glasses         &  Facial/Head Gesture Detection   & Active & Load  & Time & RC Circuit & RF &  Predicting 12 facial and head gestures with reasonably high accuracy 91.25\% &
\doitems
\item Demonstration of the technical feasibility of using capacitive sensing for facial/head gesture recognition.
\item Described performance only under laboratory conditions.
\\ 
%\cite{matthies2021capglasses}-2021   & Face      &  Glasses         &  Facial/Head Gesture Detection   & Active & Load    & Time & RC Circuit & RF &   Average accuracy of 89.6\% of 12 gestures by a user-dependent ML model. &
%\doitems
%    \item An untethered battery-powered glasses prototype to sense facial expressions and head gestures actively.
%    \item Described performance only under laboratory conditions.
%\\

\cite{luo2020eslucent}-2020   & Eyes      &  Eyelid stickers         &  Eye Blink Detection  & Active & Load  & Time & RC Circuit & statistical analysis &  An average precision
of 82.6\% and a recall of 70.4\% across all participants was achieved for blink detection. &
\doitems
    \item A novel on-skin eyelid interface for tracking eye blinking is developed.
    \item Limited users, only for those with a habit of eyelid stickers.
\\

\cite{liu2022non}-2022   & Eyes      &  Glasses         &  Eye Blink Detection  & Active &  Load  & Frequency & FDC2214 & statistical analysis &  The feasibility and robustness were demonstrated in five scenarios with an average precision of 92\% and recall of 94\% &
\doitems
    \item A non-contact, real-time eye blink detection prototype with capacitive sensing.
    \item blink detection performance degrades if the glass doesn’t match the volunteer’s head form.
\\

\cite{matthies2017earfieldsensing}-2017 & Ears        & Earbud        & Facial Expression Recognition    & Passive & Load    & Current  & Discrete component, Amplifier & J48 DT & 5 facial gestures with a precision of 90\% while sitting and 85.2\% while walking  &
\doitems
    \item A novel input method for
mobile and wearable computing using facial expressions with
electrodes placed inside the ear canal.
    \item Limited number of users(3) in data collection.
\\ 

\cite{haescher2015capwalk}-2015  & Leg & Leg- band & Walk style recognition & Active & Load & Time  &  RC Circuit & Classic machine learning & Around 87\% recognition recall of five different walking style. &
\doitems
    \item The recognition performance from leg band prototype shows best result, since it solely relies on differences in stride frequency and leg distance which is easy to observe.
    \item No practical experiments were conducted, data was from laboratory.
\\

\cite{cheng2010active}-2010  & Leg/ Neck/ Wrist & Leg/ Neck/ Wrist-band & Locomotion Detection & Active & Load & Frequency  &  Colpitts oscillator & qualitative analysis & Obvious signal differences between series of locomotion &
\doitems
    \item Described the concept, implementation, and evaluation of a new on-body capacitive sensing approach to derive activity related information.
    \item Lack of quantitative analysis.
\\

\cite{singh2015inviz}-2015  & Leg & Leg-textile & Personalized gesture recognition & Active & Load & Frequency  &  Capacitance-to-Digital Converter & Nearest Neighbor & 93\% classification accuracy of 16 hand hover and swipe postures &
\doitems
    \item Presented the design, implementation, and evaluation of a low-cost gesture recognition system for paralysis patients that uses flexible textile-based capacitive sensor arrays for movement detection..
    \item Personalized training is needed for high classification accuracy; Test on real paralysis patients is expected.
\\ 

\cite{bian2021systematic}-2021 & feet  & Shoe  & Gait partitioning, step counting & Active & load  & Frequency  &  Timer 555 & Threshold, Peak detection & Better step counting accuracy(99.4\%), and better gait partitioning (accuracy of 95.3\% and 93.7\% for stance and swing phases) than IMU &
\doitems
    \item A systematic exploration of the human body capacitance and its applications.
    \item Step counting and gait partitioning performed only indoors, outdoor performance is not clear.
\\ 

%\cite{min2018development}-2018 & feet  & Insole  & Gait monitoring & Active & load  & Frequency  &  FDC1004 & Comparative analysis & 1.77\% error rate in step counting, 1\% error rate of stride time. &
%\doitems
%    \item Proposed a system being helpful in development of gait monitoring and measurement system as smart healthcare.
%    \item Certain clothing types might impact capacitance measurements.
%\\ 

\cite{bian2021capacitive}-2021  & Wrist       & Wrist-band          & Hand Gesture Recognition & Active & Load   & Frequency  & FDC2214 & CNN & Seven hand gestures with an accuracy of 96.4\% &
\doitems
    \item Real-time on-board hand gesture recognition.
    \item Limited number of users in data collection.
\\

%\cite{grosse2012enhancing}-2012  & Wrist  & Wrist-band          & Activity Recognition & Passive & Load   & Frequency  & 555 Timer & SVM & Nine everyday-life activities with an F-score of 73.5\% &
%\doitems
%    \item Demonstrated the benefit of capacitive proximity sensor for daily activity recognition.
%    \item Large conductive electrode and no exploration of influence from electrode displacement.
%\\

%\cite{rekimoto2001gesturewrist}-2001  & Wrist       & Wrist-band          & Hand Gesture Recognition & Active & transmit & Frequency  & Not given & Not given & No numerical result is given &
%\doitems
%    \item Presented an unobtrusive capacitive gesture commanding system.
%    \item Only physical background description, missing real-life demonstration.
%\\

\cite{braun2014towards}-2014 &  Hand    & Armrest         & Hand Gesture Recognition    & Passive & Shunt   & Current  &  TMS320F2 &  SVM & Detection rate between 77.3\% and 90.9\% for touch set and free-air set between 45.5\%  and 81.8\% &
\doitems
    \item  Presented an invisible interaction device that can be seamlessly integrated into car interiors and supports touch and free-air gestures.
    \item Accuracy needs to be improved; Different sensing surface needs to be developed for cars without suitable armrests.
\\

\bottomrule
%\end{tabularx}
\end{tabular}
\end{table}
\end{landscape}
\restoregeometry
%}

%\afterpage{%  
\newgeometry{margin=2cm} % modify this if you need even more space
\begin{landscape}
\begin{table}[htbp]
 \caption{Body-part electric field applications(2)}
\label{BodyPartTable2}
%\renewcommand{\arraystretch}{1.4}
%\begin{tabularx}{\textwidth}{@{}l*{8}{C}c@{}}
\footnotesize
\begin{tabular}{ p{1.2cm} p{0.5cm} p{0.6cm} p{1.7cm}  p{0.8 cm} p{0.8cm} p{0.9cm} p{1.2cm} p{1.0cm} p{3.8cm} p{8.0cm}}
\toprule
References-Year   & Body Part & Form & Subject & Passive/ Active & Sensing Mode & Source Signal & Hardware & Algorithm & Performance & Contribution and Limitation  \\ 
\midrule

\cite{liu2020fpga}-2017 & Hand   & Off-body Patch  & Hand gesture recognition & Active & Transmit  & Frequency  &  Capacitance-Voltage convertor &  LSTM & The accuracy of 10-fold and leave-one-user-out cross-validation accuracy is respectively 97.5\% and 91.3\% &
\doitems
    \item  Proposed a method to detect driver's movement information using capacitive ECG sensors.
    \item Limited test data.
\\

\cite{wong2021multi}-2021 & Finger  & Finger-stall  & Hand gesture recognition & Active & Load  & Frequency  &  555 Timer &  SVM, KNN &  A classification rate of 16 american sign language with approximately 99\% for intra- and inter-participant data &
\doitems
    \item Presented a prototype of wearable capacitive sensor unit to capture the capacitance values from the electrodes placed on finger phalanges for hand gesture recognition. ´
    \item Power consumption; Bulky prototype.
\\

\cite{wilhelm2020perisense}-2020 & Finger   & Ring  & Multi-Finger Gesture Interaction Utilizing & Active & Load  & Frequency  &  FDC1004 &  one nearest neighbor & The leave-one-out cross-validation test using only capacitive measurements revealed an average accuracy of 88\& for eight finger gestures, 90\& for eight unistroke gestures &
\doitems
    \item Presented PeriSense, a serf-contained ring-shaped interaction device enabling multi-finger gesture interaction. 
    \item Power consumption caused by the Bluetooth module.
\\

\cite{choi2017driver}-2017 & Heart        & Seat         & Non-contact ECG Detection  & Passive & Load   & Current  &  cECG &  Independent component analysis & An averaged correlation coefficient value between the ture movement signal and ECG-decomposed movement signals was 0.77 &
\doitems
    \item  Proposed a method to detect driver's movement information using capacitive ECG sensors.
    \item Limited test data.
\\

\cite{hirsch2014hands}-2014  & Neck & Neck-band & Head-Gesture Recognition & Active & Load & Frequency  &  Colpitts oscillator & Bagging trees & Recognisation of 15 head gestures with mean accuracy is 79\%. &
\doitems
    \item Designed a textile neckband, allowing continuous unobtrusive head movement monitoring.
    \item Low accuracy for head posture monitoring.
\\

\cite{cheng2013activity}-2013  & Neck & Neck-band & Nutrition Monitoring & Active & Load & Frequency  &  Colpitts oscillator & J48 DT & The overall accuracy for all 3 subjects and 5 states (sleep, quiet, normal, active, eat) in the 1.5min windows is
84.4\%. &
\doitems
    \item Proposed an unobtrusive sensor for reliable recognisation of major meals and sleeping periods.
    \item Lack of generalization since only three subjects joined the experiment.
\\

\cite{suzuki2020mouth}-2020 & Mouth       & Mask         & Mouth Gesture Recognition    & Active & Shunt   & Frequency  &  Analog Discovery 2 &  RF & 75.4\% leave-one-session-out accuracy on five mouth gestures &
\doitems
    \item Handsfree interface recognizing non-verbal mouth gestures.
    \item Lack of leave one user out cross validation.
\\

\cite{cheng2008body}-2008  & Torso & Clothes & Control gesture Recognition & Passive & Load   & Frequency  & MC34940 & Peak Detection & Three hand control gestures with an accuracy of around 89\% &
\doitems
    \item Design of textile, multi electrode capacitive on body sensing for contact less detection of simple control gestures.
    \item Noise from shape changes of the flexible electrode and disturbances due the human body being close by.
\\

\cite{ianov2012development}-2012  & Arm & Clothes & Non-contact bioelectrical signal detection & Passive & Load   & Current  & Instrumen-tation Amplifier & Observation & Capable of correctly recording ECG and EMG under different loads &
\doitems
    \item Demonstrated the feasibility of reliable ECG and EMG recording with non-contact capacitive electrode over clothing.
    \item Long term experiments under real world conditions(surrounding electric noise, wetness of skin, etc.) are still necessary.
\\

%\cite{braun2009using}-2009 & Ambient  & Off-body patch  & Gesture Detection, etc & Passive & load  & Current  &  CY3235 &  Rule-based event & Qualitative analysis &
%\doitems
%    \item The proof-of-concept prototype based on generic hardware and simple components. 
%    \item Disturbances factors: environmental situation, setup stability, grounded conductors around, temperature and humidity.
%\\ 

%\cite{braun2013capacitive}-2013 & Ambient  & Off-body  & Hand festure recognition & Passive & load  & Current  &  CY3271 & Look-up table & By only a handful gestures, the system shows reliable basic home control &
%\doitems
%    \item Presented a generic framework for hand gesture recognition that is tailored to input devices based on arrays of capacitive proximity sensors.
%    \item Considerable sensor noise limits the gesture set.
%\\ 

\cite{erickson2018tracking}-2018 & Robot  & Off-body  & Robot assistant living & Passive & load  & Current  &  MPR121 & Curv-fitting & The robot successfully
pulled the sleeve of a hospital gown and a cardigan onto the right
arms of 10 human participants. &
\doitems
    \item Demonstrated that a simple capacitive sensor can be used to track human motion and adapt to pose estimation errors during robot-assisted dressing.
    \item Certain clothing types might impact capacitance measurements.
\\

\bottomrule
%\end{tabularx}
\end{tabular}
\end{table}
\end{landscape}

}

\restoregeometry

\noindent
sneaking, normal walking, fast walking, jogging, and walking with weight, and got a recall of around 87\% with different classic machine learning models. Compared with the prototype being deployed in the sole and around the chest, the leg deployment supplied the best result since the read mainly relies on stride frequency and leg distance, which is easily sensed by the leg band. Cheng et al. \cite{cheng2010active} also designed a capacitive leg band to sense a series of different locomotion, but with a Colpitts oscillator as the front end. By deploying the front end on other body parts, like the neck and wrist, motion patterns of the neck and wrist were demonstrated recognizable by the proposed sensing unit. In \cite{singh2015inviz}, the authors designed a capacitive electrodes array and deployed it on a textile near the leg, aiming for gesture recognition for paralysis patients. An accuracy of 93\% was achieved among 16 hand hover and swipe postures with the nearest neighbor classifier.

\subsubsection{Feet}

The movement of feet is demonstrated to be one of the critical influence factors of body capacitance \cite{bian2021systematic}, since they are the closest organ to the ground. By observing the movement of feet, a series of applications could be developed, like step counting, gait analysis, etc. In \cite{bian2021systematic}, the authors deployed a 555 timer-based capacitance monitoring front end near the feet and an electrode under the sole, a much more straightforward signal of feet movement is captured, and better accuracy of step counting and gait partitioning is achieved, compared with the traditional IMU-based solutions with the same deployment (99.4\% vs. 94.3\% with step counting, and 95.3\% / 93.7\%  vs. 93.1\% / 90.8\% with stance and swing phases duration). Besides the instrumented shoes, the insole is another medium for capacitive-based foot movement monitoring. Se et al. \cite{min2018development} used another frequency-based capacitive sensor, FDC1004 from Texas Instruments, and developed a customized insole for health status monitoring. The insole was designed with two shield layers and a sensor layer and aimed at health status monitoring by analyzing the step count and stride time. The single-channel smart insole demonstrated superior performance compared to commercial pressure-based feet monitoring systems like F-scan, composed of nearly one thousand channels. Capacitive-based gait analysis has been widely explored in the past decade \cite{zheng2017gait,min2012step,zhang2019low}, benefitting from its straightforward signal form, low-cost and uncomplicated hardware, and other properties like non-contact and non-intrusiveness. However, since the capacitive signal is mainly contributed by the proximity of the applied electrodes to the sole of the foot and ground, it is less robust than the other sensing modalities, like the resistive-based pressure sensor \cite{xu2012smart, cho2017design}. Experiments showed that different ground types could result in signal variation, especially in signal strength \cite{bian2021systematic}.

\subsubsection{Wrists}

For the deployment of a wearable assistive device, the wrist always comes first, as the wrist is, first, the most frequently used body part to reach the other positions around the body; second, the best location for holding a device without much uncomfortableness. By placing an electric field sensing band around the wrist, hand gestures could be deduced based on the principle that hand movement can disturb the electric field around and in the wrist. In \cite{bian2021capacitive}, the authors presented a capacitive-sensing wristband surrounded by four single-end electrodes for onboard hand gesture recognition. The capacitive sensing front-end is based on FDC2214 from TI, which employs an L-C tank as a sensor node. The working principle is that the L-C tank's capacitance change can be observed as a shift in the resonant frequency. The device outputs a digital value proportional to frequency, which can be converted to an equivalent capacitance value. By deploying a single convolutional hidden layer as the classifier at the sensing edge, the wristband can recognize seven hand gestures from a single user with an accuracy of 96.4\%. However, this accuracy is achieved by the data collected from a single user. Because of the wrist diversity of different subjects, the generalization of the proposed band is still challenging. A similar wristband was also introduced in \cite{rekimoto2001gesturewrist} for gesture recognition, which mainly presented the physical background and outlooked potential applications, a numerical analysis was not provided. Other works like \cite{grosse2012enhancing, reinschmidt2022realtime} were also published on the subject of capacitive-based hand activity recognition in the form of wristbands. In \cite{reinschmidt2022realtime}, a charge variation integrated sensor named with Qvar was presented for wrist-worn hand gesture recognition, and an accuracy of up to 87\% across subjects was achieved combined with IMU, where the Qvar sensor shows an enhancement by more than 10\% with respect to the IMU-alone estimations.

\subsubsection{Hands/Fingers}

Besides the wrist, hand and finger were also explored for gesture recognition in a capacitive approach. In \cite{braun2014towards, liu2020fpga}, multiple capacitive electrodes are arranged in the form of an array, and the approaching of hand and motion style of the approached hand can be sensed. In \cite{braun2014towards}, a receiver electrode is used to measure the displacement current from a transmitter electrode. When a body part enters the electric field, the field between a transmitter and receiver is interrupted. This results in a decrease in displacement current. The shunt mode sensor measures the displacement current floating from a transmitter electrode to the receiver electrode. With an SVM classifier, four touch gestures and four free-air gestures were recognized with an accuracy of 45.5\% to 90.9\%. In \cite{wong2021multi, wilhelm2020perisense, chen2023efring}, the electrodes are deployed on the fingers, aiming to sense the motion pattern of the fingers for gesture recognition. Such works utilized the frequency signal as the source of capacitive sensing. \cite{wong2021multi} reported a near-perfect accuracy with KNN on the American sign language dataset with 99\% in both intra- and inter-participant cross-validation. \cite{wilhelm2020perisense} developed the system in a ring form named with PriSense, in which the gestures of the finger-wearing ring and its adjacent fingers are sensed by measuring capacitive proximity between electrodes and human skin. An accuracy of 88\% was reported for eight gesture recognition with the nearest neighbor approach.

\subsubsection{Heart}

Non-intrusive ECG monitoring has widely explored and enjoying advantages in easy-of-use and pervasive real life scenarios \cite{arcelus2013design, wannenburg2018wireless, gao2019heart}. Essentially, ECG is a voltage signal
generated by heart activity.  Capacitive ECG sensors measures the voltage changes in the body using capacitive coupling without any direct skin contact. \cite{choi2017driver} proposes a method to monitor driver’s
movement using capacitive electrocardiogram (ECG) sensors. Of which four cECGs were placed at the left and right back of the seat respectively and the ECG was measured by the difference between one left and one right sensor signals, a conductive fabric was used to reduce a common-mode noise between left and right sensors. An averaged correlation coefficient value between the ture movement signal and ECG-decomposed movement signals was 0.77, showing that the proposed
method can give information on the movement of a driver using capacitive ECG sensors. 

\subsubsection{Neck}

Like the capacitive wristband, researchers also deployed the capacitive electrodes on the neckband for gesture recognition from the head, where the capacitive electrodes are integrated into a textile neckband. As head movement caused movement of muscles, tendons, blood vessels, and other tissue can disturb the electric field distribution around the electrodes. In \cite{hirsch2014hands}, the authors presented a hands-free gesture-controlled user interface based on active capacitive sensing, where continuous unobtrusive head movement monitoring is enabled by a capacitive neckband. A study involving 15 head gestures from 12 subjects showed an overall accuracy of 79.1\% for head gesture recognition. The authors also presented another work with a similar hardware platform, which is based on the Colpitts oscillator, to monitor the nutrition, as swallowing event also caused the motion of the neck structure. With a decision tree model, the work reported a detection accuracy of 84.4\% for five activities (sleep, quiet, normal, active, and eating). However, the data used for classification came from only three subjects, thus, the challenge of lacking generalization exists.

\subsubsection{Mouth}

As in active capacitive sensing, any body part action will cause a capacitive variation signal. Another useful capacitive gesture sensing is the month, where the signal source is the lips' action. In \cite{suzuki2020mouth}, the authors embedded the mutual-capacitance sensor in the mask to recognize five mouth gestures. A mutual-capacitance sensor features multiple intersecting electrodes. Those facing in one direction collectively serve as a transmitter delivering a sine wave, and those facing in the opposite direction as the receiver. When an element approaches the intersections, that element interacts with the electrical field, and the intersection capacitances change. A random forest classifier was trained to recognize the mouth shapes, and an average recognition accuracy of 75.4\% was reported with Leave-One-session-Out Cross-Validation, which showed the feasibility of the proposed approach for mouth-based gesture recognition and interaction.

\subsubsection{Others}

Besides the above-listed body parts that are used for capacitive sensing with the background of their conductive property, the electrodes of a capacitive sensing front-end can also be deployed on/near other body parts or in the near environment for specific applications. For example, authors of \cite{cheng2008body}  described the use of textile, multi-electrodes capacitive sensing for contactless detection of simple control gestures. The textile electrodes were sewn into the doctor’s coat and formed a linear array of electrodes. When a finger or a hand approaches the electrodes, the measured capacitance grows as the distance decreases. The sensing component is an MC34940 chip that provides seven input channels and generates a low radio frequency sine wave with nominal five volts peak-to-peak amplitude. The amplitude and phase of the sinusoidal wave at the electrode are affected by objects in proximity. A quantitative evaluation showed that high accuracy (around 89\%) could be reached for three gestures (slide-up, slide-down, and short touch). Still, unintended contact with the touchpad strip during everyday activities causes a high rate of false positives. Similar to this work, Ianov et al. \cite{ianov2012development} designed a high input impedance (level of Tega Ohm) front-end to sense the bioelectrical signal of the body (ECG, EMG, etc.) in a non-contact way when deploying the electrode on the clothes. Noncontact capacitively coupled electrodes react to electrical field variations caused by bioelectrical activity, which eliminates the need to maintain resistive contact between the skin and the electrode. This work demonstrated the feasibility of reliable EEG/EMG recording with non-contact capacitive electrodes over the clothing. 
When the capacitive sensing system is deployed off-body in the surroundings, like in the ambient or on the robot, the invasion of a particular body part will generate an efficient capacitive signal which can be used for activity inferencing. In \cite{braun2009using, braun2013capacitive}, the authors used the CY32xx series of the capacitive sensor as the front-end and deployed the sensing electrodes in the ambient for hand gesture recognition. The work of \cite{braun2013capacitive} described a method for hand tracking in three dimensions based on arrays of capacitive proximity sensors. The benefit of this setup is that the system can be hidden behind materials such as wood, wool, or plastics without limiting their functionality, making them ideal for application in ambient intelligence scenarios. \cite{erickson2018tracking} presented a dressing assistance method using capacitive proximity sensing to track the pose of a person in real-time. Compared with other dressing assistance methods like visual solutions, the capacitive approach gives direct estimates of distance with low latency, has a high signal-to-noise ratio, and has low computational requirements.

\subsection{Whole-body electric field sensing}

\begin{figure*}[h]
\centerline{\includegraphics[width=0.55\textwidth, height=6.5cm]{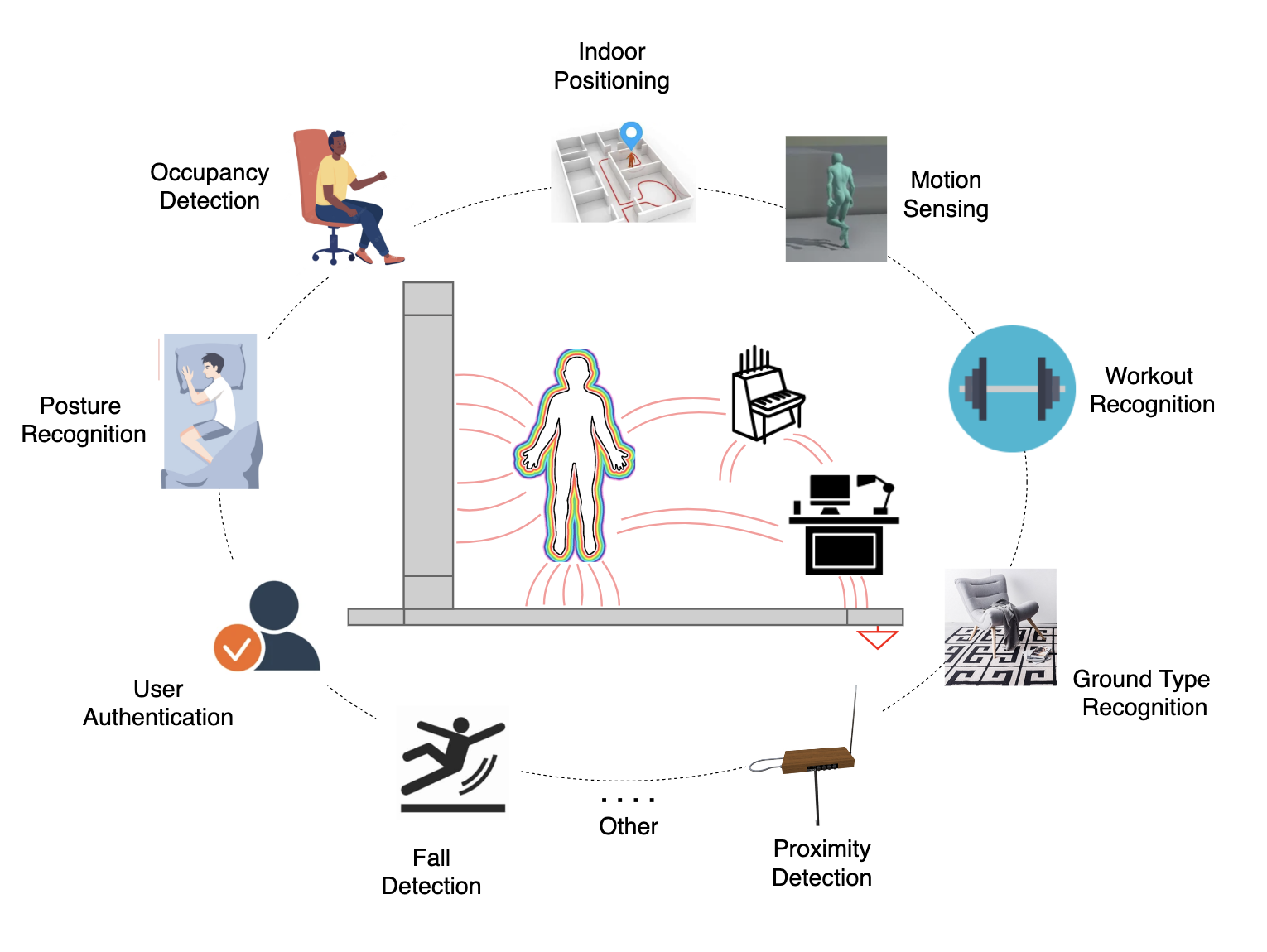}}
\caption{Applications leveraging the electric field existing/generated around the whole body}
\label{Body_Whole}
\end{figure*}

Unlike body-part electric field sensing, whole-body sensing makes use of either the existing electric field from the whole body itself in a passive way or an actively generated electric field that can be interfered with by the entire body. Especially the natural passive electric field of the body, or so-called human body capacitance ( demonstrated as around 100 pF \cite{aliau2012fast,serrano2003power,sualceanu2004measurements, fujiwara2002numerical}), as Figure \ref{Body_Whole} depicts, describing the field between the body and the surrounding (mainly the ground), with the background that the body is actually an ideal conductivity, being able to store and exchange charges. By sensing the body charge flow pattern, certain body activities could be inferred, like motion classification, indoor localization, user authentication, etc. Besides that, researchers also developed platforms that actively generate an electric field to detect the proximity and activities of a body from the signal variation caused by body invasion. The generated electric field is mainly in the AC form to resist environmental noises. Table \ref{WholeBodyTable1} and Table \ref{WholeBodyTable2} summarized the works published in the last decade when the human body is treated as either a disturbance of a radiator of an electric field, mainly sorted by the targeted subjects.

\subsubsection{Activity Recognition}

As the surrounding varies in practical scenarios, the body capacitance, namely the passive electric field between the body and the surroundings, is not a constant value. Body postures \cite{fujiwara2002numerical}, garment \cite{jonassen1998human}, skin state(moisture, etc.) \cite{goad2016ambient,egawa2002effect}, etc., are all potential influence factors to the body capacitance. The movement of a body part, like the arm and leg, will change the relative position of the body and surroundings, thus will result in a body capacitance variation. In a controlled environment, the feet' actions contribute the most to the variation of the body capacitance, as it changes the distance between the body and the ground significantly. By sensing the capacitance variation, certain body activities could be deduced. 

In \cite{cohn2012ultra}, the authors designed a customized front end composed of an amplifier and a filter to sense the charge variation caused by body actions, like waving arms and legs. The sensing approach is completely passive, relying
on the existing static electric field between the body and the environment. The designed hardware does not transmit any signal and simply measures a voltage at any single location on the body. In the user study, four types of activities were considered for activity recognition: rest, small arm movements, walking, and jogging. With six statistical features from the single-dimensional data, a KNN classifier was trained with three-fold cross-validation, and an averaged accuracy of 91.7\% was achieved for the four classes. To be noticed, the authors reported the power consumption at the sensing side with only 3.3 uW, which is so low that the power consumed by the front end is even negligible. Which makes the proposed approach ideal for continuously sampling coarse-grain body movement.

Besides the above body capacitance-based activity recognition, researchers explored more complex activities for classification. In \cite{bian2022using}, the authors explored how this novel body capacitance concept could contribute to the recognition of leg-dominated exercises, as they observed that the body capacitance has a novel property that makes it competitive and complementary to traditional wearable motion sensors(namely the inertial measurement unit): it does not require the sensor to be placed directly on the moving part of the body of which the motion needs to be tracked. To demonstrate this, they explored exercise recognition and counting of seven machine-free leg-alone exercises. The applied sensing front-end is based on the high-precision bio-potential signal sensing chip ADS1298 from TI, combined with a charge supplier and filter composed of several discrete components. By trying both traditional machine learning based on feature extraction and the deep neural network for exercise recognition, the body capacitance sensing shows significant advantages over the inertial measurement unit signals in both classification(0.89 vs. 0.78 in F-score) and counting for the leg-dominated exercises. Since the body capacitance sensing could be low-power and low-cost and ideal for wearable consumer devices for specific motion sensing, the authors also explored the edge performance with body capacitance and inertial measurement unit on AI-enabled edge devices \cite{bian2022exploring}. Besides the above research activities, the authors also collected twelve gym activity data with body capacitance signals to check the signal's recognition potential in real-life gym sessions. With the TCN blocks,  they demonstrated the feasibility of full-body workout counting and, to some extent, recognition with body capacitance, regardless of the individual habit of movement speed and scale.

\subsubsection{Indoor Localization}

Reliable indoor person location is essential for location-based services that could improve life quality and safety, especially for patients, the elderly, or any person with special needs for medical observation or accident monitoring for assistive healthcare. A wide range of sensing solutions have been explored in the past decades, like visual solutions\cite{mautz2011survey}, RF-based solutions like WIFI and Bluetooth \cite{yang2015wifi, kalbandhe2016indoor}, field-based solutions like a magnetic field \cite{bian2020social}, etc., as summarized in \cite{mendoza2019meta}. 
%The sensing technique should be tagless, passive, privacy-aware, and unobtrusive for wide adoption. 

In \cite{ramezani2016tagless}, the authors introduced a tagless indoor person localization system based on a capacitive sensing platform that utilizes a transducer operating in load mode. The sensed capacitance varies based on the distance to a nearby human body. To extend the sensing range, the authors also proposed a machine learning method as the positioning algorithm to allow the sensors to sense a person at distances suitable for localization within the area of interest. Real-life experiments where a few 555 Timer-based sensor nodes were deployed at the edge space showed that the tag-less system could track the position of a person in a 3 m × 3 m room with about 20 cm error
and better than 70\% recall and precision. 

Similarly, the authors in \cite{tang2019indoor} also deployed a few capacitive sensor nodes at room scale for indoor localization, but with the ESP charge sensor followed by a notch filter. Different from the 555 timer, which senses the capacitance variation by observing the frequency shift, the ESP charge sensor senses the capacitance variation by monitoring the charge flow on the electrode. with a high impedance input and a bias voltage, the subtle charge flow signal is perceived in the form of a voltage signal. Experimental results in two different rooms for one-week duration show the effectiveness of the proposed sensing system: occupant localization methods with room occupancy estimation accuracy of 89.03\%, and average localization error of less than 0.3 m. Besides this, the authors also described extra functionalities with their hardware. The first is respiration monitoring, with six participants involved in this experiment,  EPS sensor shows an average error rate of 4.9\% compared with ground truth. Which successfully demonstrates the capability of EPS for remote biosignal sensing. The second is participant identification, as the waveforms and frequency spectra of measured respiration signals from different people are quite diverse. With a SVM model, an accuracy of 98.3\% on six participants was achieved for .human identification.

Besides deploying the capacitive sensor node around the positioning space, researchers also explored the mat-form capacitive sensing system for indoor positioning. As described in \cite{braun2011capfloor}, a flexible, integrated solution based on affordable, open-source hardware that allows indoor localization and fall detection is presented. The system comprises sensing mats that can be placed under various types of floor covering that wirelessly transmit data to a central platform providing localization and fall detection services. The electrodes are applied in two layers insulated with each other in the mat using an insulated wire. Two wires are connected to each electrode in order to increase the spatial range of a single sensor. The achievable resolution depends on the number of sensors placed on a mat. The paper did not supply a detailed positioning performance of the design with hard numbers. Although this method could deliver high positioning precision, the installation of such a hardware system can be complex.

\subsubsection{Touch and Proximity}

As our homes are equipped with many electric appliances, like TV, refrigerator, etc., and other electric components like light switches, power lines, etc., each of them will radiate an electric field to the grounding potential level. The conductivity property of the human body will pick up those subtle electric fields and express them in the form of surface potential. The authors of \cite{cohn2011your} thus utilized the human body as an antenna to receive those electric fields that already exist in our environments and are usually treated as electric noises. A conductive pad was placed on the back neck and wired to a National Instruments USB-6216 data acquisition unit, which sampled the pad potential at 400 kS/s. A bias voltage to the local ground signal on the data acquisition unit was applied to the pad potential with a 10 MOhm resistor. The signal was digitized at 16-bit resolution and streamed to disk on an attached
laptop for subsequent processing. By observing the properties of the electric noise picked up by the human body in a series of experiments, the authors concluded that this new sensing modality could offer capabilities of robust classification of gestures with a fingerprinting approach, such as the position of discrete touches around light switches, the particular light switch being touched, which appliances are touched, differentiation between hands, as well as continuous proximity of hand to the switch, etc. Meanwhile, the authors also discussed the limitation of the proposed sensing modality, like the generalizability of the observed noise signals, as the electrical noise is a side effect of the power line infrastructure, there are no simple predictive models to infer what the signal will look like at different locations; And the state issue of the home, as the classification works well when the home is in the same state as it was during training, but a large change in the home state (i.e., turning on the air conditioning, or all lights in the home) will cause the classification accuracy to drop. The authors also outlooked the potential solutions for such limitations. Overall, the work provided a novel insight for developing assistive and interactive home living styles at very low cost, as the high cost of instrumenting environments limits the ubiquity of even mature technologies for smart-home.

\subsubsection{Ground Type Recognition}

Ground types are usually distinct in the everyday living space, especially in an office building. Recognizing the ground types a person is standing on and switching between could supply an additional signal for assisted living, like indoor positioning. Unfortunately, there are seldom sensing techniques that could recognize the ground types in a wearable way. In \cite{bian2021systematic}, the authors found that the body capacitance(both its absolute and variation values) while walking will have different patterns when the individual is walking on various ground types. This is reasonable since the body capacitance describes the electric field between the body charge plate and the ground charge plate. An action of feet, like lifting and dropping the feet off and to the ground, will change the distance between the two charge plates; Similarly, different ground types mean different dielectrics, thus also causing different values of body capacitance. This observation was obtained by attaching an electrode under the sole and insulating the electrode from the ground. A 555 Timer is used to interpret the capacitance variation into the frequency values. The study collected 28 sessions of body capacitance variation data from seven volunteers on six types of ground (textile, carpet, indoor concrete, stairway concrete, wood, and outdoor cement brick). Each volunteer walked indoors to outdoors and back with the same path four times. With a Random Forest model, a combined F-score of 0.63 is achieved in the six ground types recognition. The accuracy is robust as it was achieved with both leaving one session out and leaving one user out in training. Besides this, the authors also noticed that the body electric field strength variation in the exit stairway while walking from indoors to outdoors is significantly distinct from the mode while walking back. This observation is uniform in all sessions among all users. Thus a potential conclusion can be made that the body's electric field strength relates not only to where the body is but also to where the body used to be.

\subsubsection{User Authentication}

User authentication is a process that verifies the identity of the
the user of a computing device or an online service. It plays an important role and has a wide range of practical usage scenarios in everyday life when interacting with the digital world, like in the field of security \cite{clarke2007advanced}, automation \cite{park2009smart}, and entertainment \cite{angulo2011exploring}. With the popularization of mobile devices like tablets, the act of touching the screen happens more often. But who is touching the screen is not easily differentiable. In \cite{harrison2012capacitive}, the authors present a non-instrumented approach for user authentication when a screen is being touched. The novel sensing approach is based on swept frequency capacitive sensing and can be named capacitive fingerprinting, allowing touchscreens to not only report the touch location but also identify the user that makes the touch action. The background is that the human body is unique regarding biological and anatomical factors, and the wearing also alters how a user is grounded. The swept frequency capacitive sensing estimates the impedance profiles of users at different frequencies. The impedance profiles include the resistance profile and capacitance profile, which implies the body's electric field property. In the study, the chip AD5932 is embedded into a tablet as the sensing unit, without any instruments on the body. An average accuracy of 97.3\% is achieved with an SVM classifier,  which shows a considerable promise of

\afterpage{% 
\newgeometry{margin=2.0cm, left=3.5cm} % modify this if you need even more space
\begin{landscape}
\begin{table}[]
 \caption{Whole-Body electric field applications(1)}
\label{WholeBodyTable1}
\renewcommand{\arraystretch}{1.4}
%\begin{tabularx}{\textwidth}{@{}l*{8}{C}c@{}}
\footnotesize
\begin{tabular}{ p{1.2cm} p{1.5cm} p{1.5cm}  p{1.0 cm} p{1.0cm} p{1.2cm} p{1.2cm} p{3.8cm} p{8.0cm}}
\toprule
References-Year & Subject & Passive/ Active & Sensing Mode & Source Signal & Hardware & Algorithm & Performance & Contribution and Limitation  \\ 
\midrule

\cite{cohn2012ultra}-2012 & Body motion sensing & Passive & Load & Current & General amplifier & KNN & Nearly 92\% classification accuracy of rest, small arm movements, walking, and jogging. &
%\vspace{-\topsep}
\doitems 
\item Presented an ultra-low-power method for passively sensing body motion using static electric fields by measuring the voltage at any single location on the body.
\item Charge changes caused by surroundings confused the system.
\\

\cite{bian2022using}-2022 & Leg-exercise recognition & Passive & Load & Current & ADS1298 and a front end & RF and DNN & Body capacitance performs better than IMU in classification(0.89 vs 0.78 F-score) and counting. &
\doitems
    \item Demonstration of the advantages of body capacitance signal over the IMU signal for leg-exercise recognition.
    \item Experimented only in an office area.
\\

%\cite{arshad2017capacitive2}-2017 & Human motion sensing & Active & Load & Frequency & 555 Timer & Curv-fitting & Qualitative analysis. &
%\vspace{-\topsep}
%\doitems 
%\item Proposed a cheap, non-invasive capacitive proximity-based floor sensing system to monitor the motion of a human body.
%\item No quantitative analysis is given.
%\\

\cite{ramezani2016tagless}-2016 & Indoor Localization & Active & Load & Frequency & 555 Timer & KNN, NB, SVM &   Track the position of a person in a 3 m × 3 m room with about 20 cm error &
\doitems
    \item  A tag-less passive localization capacitive system that can track the position of a person in a 3 m × 3 m room with about 20 cm error.
    \item Experimented only in a laboratory area.
\\

\cite{tang2019indoor}-2019 & Indoor occupancy Awareness/ Localization & passive & load & current & EPS & Maximum likelihood & human identification with an accuracy of 98.3\%, occupant localization accuracy of 89.03\% with average error less than 0.3 m. &
%\vspace{-\topsep}
\doitems 
\item Introduced a sparse low-power sensor network using passive electric potential sensors for crowd-aware smart buildings.
\item Environmental dependency results in location degradation.
\\

\cite{braun2011capfloor}-2011 & Indoor positioning & Active & Load & Time & TMS320F2 & weighted-average & Localization precision on the fully equipped floors was in the range of 50 cm. &
%\vspace{-\topsep}
\doitems 
\item Proposed a flexible, unobtrusive indoor positioning solution based on affordable, open-source hardware.
\item Limited covering area; deployment complexity.
\\

\cite{cohn2011your}-2011 & Touch/ Proximity & Active & Load & Frequency &  USB-6216 and a front end & Fingerpr-inting &   Location of the user in the home with 99\% accuracy. Position around a light
switch on a wall With 87\% accuracy. &
\doitems
    \item Demonstrated the possibility to robustly recognize discrete touched locations on an un-instrumented home wall.
    \item Lack of generalizability, calibration and training are needed for gestures in a certain home.
\\

%\cite{bian2019passive}-2019 & Workouts Recognition & Passive & Load & Current &  ADS1298 and a simple front end & TCN &  Average workouts counting accuracy of 91\% and recognition f-score of 63\%. &
%\doitems
%    \item Demonstrated the the feasibility of full body workouts counting and to some extent recognition with body capacitance, regardless of personal habit of movement speed and scale.
%    \item The recognition accuracy is low.
%\\ 

\cite{bian2021systematic}-2021 & Ground Type Recognition & Active & Load & Frequency &  555 Timer & RF &  A combined F-score of 0.63 is achieved for six ground type recognition with both leave one session and one user out. &
\doitems
    \item A systematic study of the human body capacitance, and a series of user scenarios exploration.
    \item Lack of portablity of the hardware since the body needs to be enclosed to the circuit from head to feet.
\\ 

\cite{harrison2012capacitive}-2012 & User Authentication & Active & Transmit & Frequency &  AD5932 & SVM &  Average accuracy of 97.3\% with 11 volunteers. &
\doitems
    \item Demonstrated the considerable promise of Swept Frequency Capacitive Sensing for user authentication with single touch.
    \item Lack of persistence considering environmental variation.
\\ 

\cite{sato2012touche}-2012 & Touch Interaction & Active & Transmit & Frequency &  AD5932 & SVM &  Recogntion accuracy of 95\% for 4 grasp gesture, 92.6\% for 7 body gesture, etc., with 12 volunteers. &
\doitems
    \item Demonstrated that multi-frequency capacitive sensing is valuable and opens new and exciting opportunities in HCI.
    \item The sensing technique is sensitive to variations in users’ anatomy.
\\

\bottomrule
%\end{tabularx}
\end{tabular}
\end{table}
\end{landscape}
\restoregeometry
%}

%\afterpage{% 
%\newgeometry{margin=2cm} % modify this if you need even more space
\newgeometry{margin=2cm, left=3.5cm} % modify this if you need even more space

\begin{landscape}
\begin{table}[]
 \caption{Whole-Body electric field applications(2)}
\label{WholeBodyTable2}
\renewcommand{\arraystretch}{1.4}
%\begin{tabularx}{\textwidth}{@{}l*{8}{C}c@{}}
\footnotesize
\begin{tabular}{ p{1.2cm} p{1.5cm} p{1.5cm}  p{1.0 cm} p{1.0cm} p{1.2cm} p{1.2cm} p{3.8cm} p{8.0cm}}
\toprule
References-Year & Subject & Passive/ Active & Sensing Mode & Source Signal & Hardware & Algorithm & Performance & Contribution and Limitation  \\ 
\midrule

\cite{takiguchi2007human}-2007 & Body Detection / Step Counting & Passive & load & Current &  Amplification circuit & Peak Detection &   Detecting the number of steps at an accuracy of approximately 99.4\% was shown. &
\doitems
    \item Demonstration the feasibility of walking detection remotely with simple electrode based on electrification of human body.
    \item Detection levels and polarities is not constant depending combination of footwear and floor materials.
\\

\cite{cohn2012humantenna}-2012 & Body Gesture Recognition / Location Classification & Active & Load & Frequency &  USB-6216 and a front end & SVM &  Robust gesture recognition with an average accuracy of 93\% across 12 whole-body gestures  &
\doitems
    \item Whole body gesture recognition with no instrumentation to the environment and minimal to the user.
    \item There is significant variation in ability to classify gestures at different locations.
\\ 

\cite{grosse2013opencapsense}-2013 & Proximity-based interaction & Active & Load/ shunt & Time/ Current &  TMS320F2 & Not Given & Wide application: smart couch, gesture recognition, fall detection, wearable device &
%\vspace{-\topsep}
\doitems 
\item Designed a highly flexible open-source toolkit that enables researchers to implement new types of pervasive user interfaces with low effort.
\item Costly; Influence of environment effects in targeted usage scenarios.
\\ 

\cite{braun2015capseat}-2015 & Automotive Activity Recognition & Active & Load & Time &  TMS320F2 & SVM & Achieved classification precision between 95\% and 100\% for nine static postures and dynamic events. &
%\vspace{-\topsep}
\doitems 
\item Evaluation of capacitive proximity sensing based seat for automotive activity recognition.
\item High cost(hundreds dollars); Influence of environment effects, foreign conductive objects, and different driver body type.
\\

\cite{wimmer2011capacitive}-2011 & Gaming Controller Input & Active & Transmit & Frequency &  555 Timer & Questionnaire analysis &  Using capacitive sensors can add an additional layer of complexity to game. &
%\vspace{-\topsep}
\doitems 
\item Evaluation of capacitive sensors as input modalities for different computer games. 
\item The control is easy to understand but hard to master.
\\

\cite{george2010combined}-2010 & Seat occupancy detection & Active & Shunt & Frequency & Customized \cite{zangl2008design} & Rule-based & Structured observation &
%\vspace{-\topsep}
\doitems 
\item Presented a novel combined inductive–capacitive sensor and its application to reliable seat occupancy sensing.
\item Quantitative analysis is needed to verify the feasibility.
\\

\cite{zeeman2013capacitive}-2013 & Seat occupancy detection & Active & Load & Time & Arduino UNO & Statistic analysis & The relationship of charging time between unoccupied and occupied is approximately double. &
%\vspace{-\topsep}
\doitems 
\item Presented a simple and low-cost capacitive sensor system that is ideal for occupancy detection in multiple-seat vehicles.
\item No investigation of artifacts like conductive objects on the seats.
\\

\cite{arshad2017capacitive}-2017 & Tracking and Fall detection & Passive & Load & Frequency &  Not given & Observation & Obvious signal when people contact or fall down on the instrumented floor. &
%\vspace{-\topsep}
\doitems 
\item Reported the preliminary results on a capacitive-based sensing system for elderly tracking and fall detection.
\item No quantitative analysis is conducted.
\\

%\cite{takano2017non}-2017 & ECG and respiration detection & Passive & Load & Current & Customised front-end & Curv-fitting & The sensitivity and accuracy of R/T waves were over 92\%; The accuracy of chest and abdominal RM was 70.9\% and 74.0\%. &
%\vspace{-\topsep}
%\doitems 
%\item Developed a system for simultaneous measurement of ECG and RM signals in adults without direct skin contact.
%\item Low accuracy on respiratory movements.
%\\

\cite{zhang2018wall++}-2018 & Interactive and Context-Aware Sensing & Active & Shunt & Frequency & ADS5930 & SMO classifier & 92.0\% accuracy of correct pose inference; 97.7\%  of correct touch events. 99.8\% of correct hover actions. &
%\vspace{-\topsep}
\doitems 
\item Presented a low-cost sensing approach that allows walls to become a smart infrastructure, enhancing rooms with sensing and interactivity.
\item Deployment complexity.
\\

\bottomrule
%\end{tabularx}
\end{tabular}
\end{table}
\end{landscape}
\restoregeometry

}

\restoregeometry

\noindent the described approach in user authentication by just touching the screen as usual. However, some limitations still exist, challenging the technique in real-life deployment. For example, the approach can differentiate only among a small set of concurrent users, and users can only touch sequentially instead of simultaneously. 

\subsubsection{Touch interaction}

As used for user authentication, the same technique, swept frequency capacitive sensing, could be deployed, first, on-body to recognize hand and body postures; second, into off-body universal items for a ubiquitous touch interface, empowering them the awareness of being touched or even recognizing the touch gesture. The work from Munehiko et al. \cite{sato2012touche} describes this topic. The main point of the work is to monitor the response to capacitive human touch over a range of frequencies. Since objects excited by an electrical signal respond differently at different frequencies, the changes in the return signal will also be frequency-dependent. This signal was used for contextual information extraction and enhancing the potential applications based on touch interactions. 
A few innovative example applications were described in the work. For example, the intelligent touch interaction on everyday objects, body postures tracking with a table, hand gesture recognition in the liquid environment, etc. Those evaluations showed a high and robust detection or recognition accuracy and demonstrated the utility of the technology. Different from \cite{cohn2011your}, the touch detection system described in this work needs to instrument the objects first, which introduces some complexity in practical deployment but provides a more robust touch detection that will not be inferred by environmental electric field variation, like the variation caused by home power and large appliance power on-off switch.

\subsubsection{Body Detection / Step Counting}

Body detection in the home environment plays an important role in smart homes and elderly care. While visual solutions supply direct perceiving ability, it suffers from privacy and blind corner issues. Recent RF-based solutions like WIFI \cite{wu2015non} and mmWave \cite{gu2019mmsense} show great potential with impressive reliability. Before those solutions, researchers also explored body detection with the body-area electric field. In \cite{takiguchi2007human}, the authors designed a customized front-end that can perceive the charge variation on an electrode caused by human body motion in a home environment. The authors thought that if the field intensity around the body is sufficiently high, a new method for detecting the human body that remotely detects component parts inherent in bipedal locomotion within a room based on such intensity would be possible. The approach's feasibility was successfully demonstrated in a lab-controlled experiment with a simple, remotely installed electrode followed by a few low-cost electric components. To be noticed, although the solution is based on the passive electric field of the body, the other passive electric field, such as the ones radiated from an appliance or the like, whose frequency bands are different, thus will not cause noise.

Meanwhile, from the detected body's walking signal, both feet' cadence components can be used for step counting. In the experiment, the walking of ten subjects, including infants, was measured, and the result shows that the approach is capable of detecting the number of steps at an accuracy of approximately 99.4\%. Although detecting body and counting steps with the proposed method is novel and features low cost for instrumentation, it is not perfect. For example, when multiple bodies walk simultaneously, the electrical field shows a waveform in which multiple walking components are synthesized; And a body with bare and sweet feet results in a less efficient charge variation signal. How to address those limitations is not provided by the authors, as well as by other researchers who have the same interest.

\subsubsection{Body Gesture Recognition / Location Classification}

Continuing the work described in \cite{cohn2011your} where the authors measured and digitized the voltages picked up by the human body for touch detection and localization, with mostly the same hardware platform, the authors explored full body gesture recognition and body location classification utilizing the principle that the body-area electric field caused by the electric field radiated from power lines and household appliances \cite{cohn2012humantenna} contains contextual information. Traditional methods for whole-body gesture recognition are primarily based on computer vision \cite{joseph2010framework} and inertial sensors \cite{malleson2017real}. Despite the high accuracy, vision-based approaches are generally limited in their field of view and suffer from occlusion problems. Inertial sensing approaches locate all of the sensings on the body, which is cumbersome for users with a long-term on-body sensor deployment. Recent novel RF-based solutions like WIFI \cite{zuo2021new} and mmWave \cite{li2020capturing} enjoy the advantages in reliability and accuracy but still fail in cost when compared with the passive electric field sensing approach described in \cite{cohn2012humantenna}, where the hardware system has the potential to drop to several dollars since its nothing more than a sensing front end with several simple electric components. And also the computing load of the proposed method is much lighter as the signal to be processed is only a single channel one dimensional signal. Different from \cite{bian2022using}, where the hardware platform captures the potential signal mainly caused by body motion, in this work, the hardware platform captures the potential signal more related to the home-wide environmental electric field. To test the system, the authors performed an experiment in which eight participants in different locations in eight different homes conducted twelve full-body gestures while wearing the proposed device. By performing offline segmentation and feature extraction, the SVM classifier gave an impressive accuracy of 92.7\% across participants and homes with cross-validation. Besides gesture recognition, location classification performance was also explored. With data from three participants in two homes with eight locations each and similar offline data classification, an accuracy of nearly 100\% was achieved. The authors finally also explored the real-time online performance of gesture recognization and location classification (including real-time data capturing, segmentation, and feature extraction) and obtained the classification accuracies of what was expected using the offline system. The limitation of using the body-area electric field for full-body gesture recognition described in this work is that the electrical state of the home, like turning on and off appliances and lights, will degenerate the performance reported.

\subsubsection{Proximity-based interaction}

As the interaction interface between human and computer devices that are able to determine gestures, body movements, and environmental changes at typical distances, different sensing modalities feature different profits. With the background that the human body is, on one side, conductive and able to interfere with an existing electric field when approaching, on another side, having a certain amount of charge on the surface, thus radiating a subtle electric field,
body-area electric field-based or so-called capacitive proximity sensors can be used to implement a variety of expressive input devices. As such sensors are small, robust, flexible, and can be embedded into the environment easily or worn on the body comfortably for whole-body interaction.

In \cite{grosse2013opencapsense}, the authors shared a capacitive proximity sensing platform called OpenCapSense with eight channels on top of TMS320F2. The design is able to work in both load and shunt modes, depending on the targeted application and hardware components configuration. The loading mode sensor is
based on a timer configuration called astable multivibration. The timer controls the charging and discharging cycles of the capacitor that is created by the sensing electrode and the surrounding environment. In shunt mode, a receiver electrode is used to measure the displacement current from a transmitter electrode. For example, when a human body part enters the electric field, the field between a transmitter and receiver is interrupted. This results in a decrease of displacement current and, thus, a decreasing capacitance between the transmitter and receiver. A series of example applications were explored to demonstrate its capability. Smart couch, by applying eight loading mode electrodes under the upholstery of the couch, the instrumented couch could classify 10 postures with an accuracy of more than 97.5\% with data from 18 subjects; Proximity-based hand gesture recognition, by using multiple OpenCapSense boards configured as a sensing array and a single accelerometer, the prototype could control typical multimedia applications with selection indicated by hand position and actions triggered by different knocking events that are registered by the accelerometer; Wearable devices, by integrating the electrodes unobtrusively in a wristband, it can measure the distance to objects nearby; Fall detection, by placing eight electrodes under an ordinary carpet, the smart carpet can detect a fall with a very high update rate of approximately 104 Hz and have the measurement interval down to 1.2 ms; Automotive activity recognition, as described in \cite{braun2015capseat}, by deploying two boards (sixteen electrodes) on the seat, a wide range of physiological parameters about the driver could be measured, and an advanced driver assistance system was proposed that could aid in seat adjustment, propose correct posture during the drive, and inform on signs of potential drowsiness. Similar works have also been done in \cite{george2010combined, zeeman2013capacitive} for seat occupancy detection, and in \cite{arshad2017capacitive} for fall detection, with different hardware platforms but the same principle. 

Another interesting application based on capactivie proximity sensing is the gaming controller. The authors in \cite{wimmer2011capacitive} explored the gaming performance that takes the body movement as input through capacitive proximity sensors built with the 555 Timer. 
One of the games is that the user controls a ball by walking on top of a metal plate that acts as the sensor antenna. The sensor readings allow for determining the pace and amplitude of the stride. Another game is that users play a jump-and-run video game using capacitive sensors. They could move a penguin left or right by approaching one of the two antennas mounted in front of the user. When the user jumped, the penguin would jump accordingly. Because of the drift in the reading, an adaptive moving average filter was adapted for noise and drift cancellation. Overall, these studies provide evidence that game controllers using capacitive sensors can add an additional layer of complexity to the games. The authors summarized three reasons for using capacitive sensors to sense the whole body action as the input of a game. Firstly, the hardware is simple and small and can be hidden almost everywhere. Secondly, the continuous sensor output is one-dimensional, which can be directly used as an input parameter for a game. Thirdly, capacitive sensors generate shallow complexity data, allowing for a straightforward mapping from sensor output to the input event. Users can quickly grasp the correlation between their movements and the system’s reaction. Thus the barrier to entry of such an interface can be very low. 

In \cite{zhang2018wall++}, the authors designed Wall++, a low-cost room-scale sensing approach that allows walls to become a smart infrastructure for interaction. By sensing the proximity of the body to the electrodes installed in the wall surface, the design can track users’ touch and gestures, as well as estimate body pose if they are close. Besides this, by capturing airborne electromagnetic noise, the design can also detect what appliances are active and
where they are located.

Capacitive proximity sensors have a significantly lower impact on a user’s perceived and actual privacy, compared to, for example, optical tracking methods. They are energy-efficient and can be deployed unobtrusively underneath furniture, carpets, or walls. The sensed data can be processed with computationally cheap algorithms. However, the drawbacks are limited resolution and error-proneness in environments with many conductive objects or electrical devices that affect electric fields.

\begin{figure*}[!t]
\centerline{\includegraphics[width=0.55\textwidth, height=7.0cm]{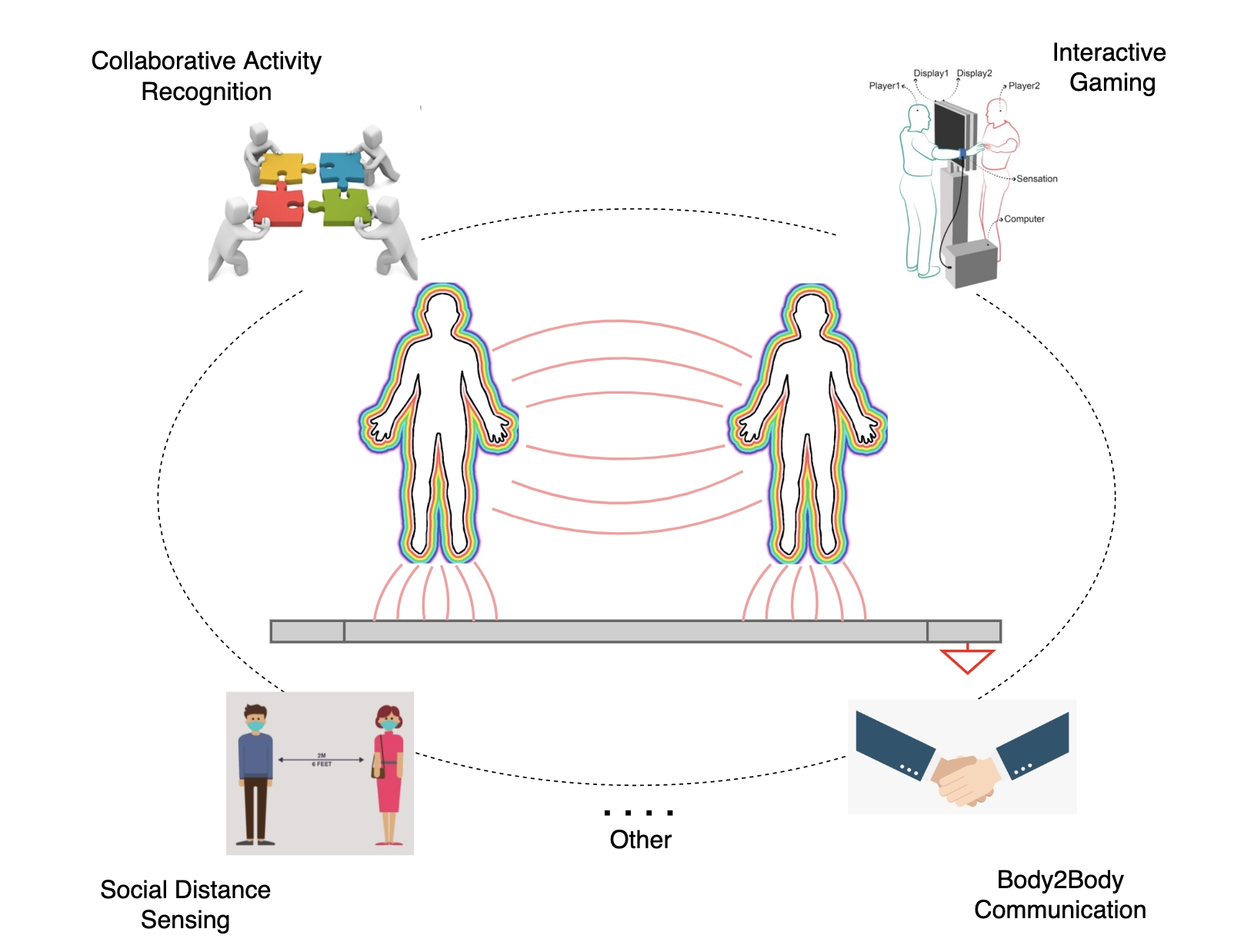}}
\caption{Applications leveraging the electric field existing/generated between different bodies}
\label{Body_Body}
\end{figure*}

\subsection{Body-to-Body electric field sensing}

As the human body radiates a slight electric field and itself is an ideal conductive object, it is interesting to explore applications in the case of multiple bodies when they affect each other's radiated electric fields and when an electric signal being transmitted across coupled or connected bodies. This subsection will present recent works on this topic, as listed in Table \ref{BodyToBodyTable}, sorted by the targeted applications. 

\subsubsection{Collaborative activity detection}

A single human body together with the surroundings, as the two conductive plates, forms a natural capacitor. When multiple bodies closely coupled, extra capacitors are formed between the bodies. Along with the movement of the bodies, those body-to-body capacitance or electric fields will be in a dynamic state. This mutual effect could be used for collaborative activity monitoring, supplying a low-cost, low-power consumption, and wearable solution, other than the other costly solutions like visual based ones \cite{li2020multi}. 

In \cite{bian2019wrist}, the authors shows how a wrist-worn charge-based body capacitance sensor signal looks like when two volunteers were collaboratively moving everyday conductive object. A signal plot of two volunteers walking with and without holding the same chair shows significantly different signals in the form of amplitude, which is a major factor for detecting collaborative work. Besides the on-body capacitive sensors, the same sensors were also deployed into the environment, being able to sense the body motions like walking by and actions like touching. Although there is no quantitative analysis, the work shows the potential of collaborative work monitoring by sensing the variation of the near-body electric fields.

Continued with the above work, the authors planned a collaborative physical work including both independent and joint activities of each worker, building a TV-Wall, as described in \cite{bian2022contribution}. Twelve participants(ten male,
two female) were divided into four groups, each of them carried some large TV screens alone or jointly from the store room to a task operating spot, assembled and disassembled a high TV-Wall, and carried them back alone or jointly. Finally,  39 sessions of valid data altogether were collected and labeled manually assisted by the ground truth from cameras. Each session contains around one hour’s motion signals from both body capacitance and accelerometer.
To show that recognizing activities across groups is learnable, a leave-one-group-out procedure was employed, where, for each fold, the test set contains all days of one group, while the training set contains all days for the remaining groups. With a logistic regression classifier for activity classes that challenge the IMU, such as carrying an object alone, carrying it jointly with another agent, or carrying nothing just walking, the result shows that the collaboration between users was detected with 0.69 F-score when receiving data from a single user and 0.78 when receiving data from both users. And the capacitive sensor can improve the recognition of collaborative activities with an F-score over a single wrist accelerometer approach by 16\%.

\subsubsection{Social Distance Monitoring}

Social distancing has been widely explored in the past three years during the COVID-19 pandemic, aiming to track the contact of a contagious patient: anyone who has been within this distance of the contagious person for a significant amount of time could potentially have been infected. An efficient social distance monitoring or alerting system can significantly decrease the risk of infection. Different sensing solutions have been proposed in the literature, like Bluetooth \cite{kumar2021social}, environmental sound \cite{bahle2021using}, ultrasound \cite{malik2020social}, UWB \cite{reddy2020social}, magnetic field \cite{ bian2020wearable}, etc. Bluetooth solution outperforms as it can be enabled by everyday used smartphones and has been widely adopted in COVID tracking Apps. However, it still faces the challenge of accuracy \cite{leith2020measurement}. In \cite{bian2022human}, the authors designed ProxiBand, a wristband that can automatically detect the intruder and alert the host by a motor-vibrating signal from the band. The background is that the intrusion of a second body will variate the passive electric field density of the first body. Thus by sensing the body electric field variation, the intrusion could be potentially deduced. The indoor experiment showed that the detection distance is ideal for the required 1.5 meters of social distancing. The authors evaluated the ProxiBand in two practical scenarios: 

\afterpage{% 

\newgeometry{margin=2cm, left=3.5cm} % modify this if you need even more space

\begin{landscape}
\begin{table}[H]
 \caption{Body-to-body electric field applications}
\label{BodyToBodyTable}
\renewcommand{\arraystretch}{1.4}
%\begin{tabularx}{\textwidth}{@{}l*{8}{C}c@{}}
\footnotesize
\begin{tabular}{ p{1.2cm} p{1.5cm} p{1.5cm}  p{1.0 cm} p{1.0cm} p{1.2cm} p{1.2cm} p{3.8cm} p{8.0cm}}
\toprule
References-Year & Subject & Passive/ Active & Sensing Mode & Source Signal & Hardware & Algorithm & Performance & Contribution and Limitation  \\ 
\midrule

\cite{bian2019wrist}-2019 & Collaborative Activity Detection & Passive & load & Current & ADS1298 and a front end & Peak Signal Detection & One-shot collaborative signal analysis. &
\doitems
    \item Presented the feasibility of collaborative work monitoring by sensing the variation of the near field electric field. 
    \item Lack of quantitative exploration.
\\

\cite{bian2022contribution}-2022 & Collaborative Activity Detection & Passive & load & Current & ADS1298 and a front end &  logistic regression & The capacitive sensor can
improve the recognition of collaborative activities with an F-score over a single wrist accelerometer
approach by 16\%. &
\doitems
    \item Demonstrated the contribution of body capacitance signal in the collaborative activity recognition task. 
    \item Single user scenario, lack of verification for generalization.
\\

\cite{bian2022human}-2022 & Social Distancing & Passive & Load & Current & ADS1298 and a front end & Threshold based proximtiy event detection & A true positive alert rate of 74.3\% when tests was performed in an indoor environment and only 46\% in an outdoor environment. &
\doitems
    \item Shows the potential of the passive body-area electric field for social distance monitoring.
    \item Low accuracy and robustness caused by environment instability.
\\

\cite{staudt_pascal_2022_6798242}-2022 & Sonification of Proximity and Touch & Active & Transmit & Frequency & Teensy and Amplifiers & quasi-quantitative analysis & Proximity measurement up to approximately 4m &
\doitems
    \item Designed a digital system that mediates between bodily movements and musical sounds, allowing for a closed-loop auditory interaction between two or more people based on electro-quasistatic coupling.
    \item Lack of robustness, human bodies do not only couple with each other and the ground, but also with the environment.
\\ 

\cite{rizzonelli2022fostering}-2022 & Social Interaction & Active & Transmit & Frequency & Teensy and Amplifiers & Structured observation & Social behaviours are solidly present during the usage of proposed system &
\doitems
    \item Designed a system that sonifies motor behaviour in real time to emerge the social behaviour through sound synthesis.
    \item Lack of solid statistical analysis.
\\ 

\cite{canat2016sensation}-2016 & Collaborative physical games & Active & Transmit & Swept Frequency & ADS9850 & Questionnaire & The overall user experience is improved when using the proposed "Sensation" system &
\doitems
    \item Demonstrated that physical contact between players may facilitate the empathy between players and increase their connectedness to each other.
    \item Capacitive effect due to interaction varies from person to person. Therefore, the Sensation can work after an individual calibration.
\\

\bottomrule
%\end{tabularx}
\end{tabular}
\end{table}
\end{landscape}
\restoregeometry
}

\restoregeometry

\noindent a social place in a building and outside a coffee shop. The results show that the Proxiband could give a true positive of 74.3\% indoors and 46\% outdoors, indicating that the prototype is feasible in the indoor environment. In the outdoor environment, it performs not as expected, as the body electric field will be much weaker than in the indoor environment where a rich amount of powered wires and appliances exiting. Although this preliminary study didn't conclude with a reliable demonstration, it shows the novelty of the body electric field for supplying new sensing approaches, as it still needs to be widely explored.

\subsubsection{Sonification of Proximity and Touch}

Unlike the above work that uses the passive body static electric field for social distancing (which signal is so tender that it is easily interfered with by environmental variations), in \cite{staudt_pascal_2022_6798242}, the authors deployed pairs of transmitter and receiver on multiple bodies, making the body itself as a part of a sensor system that works regardless of the location of the measurement and orientation of the body in the space. Benefitting from the inter-body coupling, the transmitted electric field from the transmitter can be picked up by the receiver when the two bodies are well coupled. The received signal is then used to deduce the distance information and generate the sound. The design allows for a closed-loop auditory interaction between two or more people in a physical environment. 
The evaluation shows that the sensor system can measure proximity up to approximately four meters. Above this range, noise starts to shadow the measurement, which is, therefore, covering the relevant interpersonal zones: intimate, personal, and social. A quantitative comparison is not presented in the work but will be in the subsequent user study, as the authors stated.
To be noticed, sContinued with the above work, the authors planned a collaborative physical work including both independent and joint activities of each worker, building a TV-Wall, as described in \cite{bian2022contribution}. Twelve participants(ten male,
two female) were divided into four groups, each of them carried some large TV screens alone or jointly from the store room to a task operating spot, assembled and disassembled a high TV-Wall, and carried them back alone or jointly. Finally,  39 sessions of valid data altogether were collected and labeled manually assisted by the ground truth from cameras. Each session contains around one hour’s motion signals from both body capacitance and accelerometer.
To show that recognizing activities across groups is learnable, a leave-one-group-out procedure was employed, where, for each fold, the test set contains all days of one group, while the training set contains all days for the remaining groups. With a logistic regression classifier for activity classes that challenge the IMU, such as carrying an object alone, carrying it jointly with another agent, or carrying nothing just walking, the result shows that the collaboration between users was detected with 0.69 F-score when receiving data from a single user and 0.78 when receiving data from both users. And the capacitive sensor can improve the recognition of collaborative activities with an F-score over a single wrist accelerometer approach by 16\%.

\subsubsection{Social Distance Monitoring}

Social distancing has been widely explored in the past three years during the COVID-19 pandemic, aiming to track the contact of a contagious patient: anyone who has been within this distance of the contagious person for a significant amount of time could potentially have been infected. An efficient social distance monitoring or alerting system can significantly decrease the risk of infection. Different sensing solutions have been proposed in the literature, like Bluetooth \cite{kumar2021social}, environmental sound \cite{bahle2021using}, ultrasound \cite{malik2020social}, UWB \cite{reddy2020social}, magnetic field \cite{ bian2020wearable}, etc. Bluetooth solution outperforms as it can be enabled by everyday used smartphones and has been widely adopted in COVID tracking Apps. However, it still faces the challenge of accuracy \cite{leith2020measurement}. In \cite{bian2022human}, the authors designed ProxiBand, a wristband that can automatically detect the intruder and alert the host by a motor-vibrating signal from the band. The background is that the intrusion of a second body will variate the passive electric field density of the first body. Thus by sensing the body electric field variation, the intrusion could be potentially deduced. The indoor experiment showed that the detection distance is ideal for the required 1.5 meters of social distancing. The authors evaluated the ProxiBand in two practical scenarios: a social place in a building and outside a coffee shop. The results show that the Proxiband could give a true positive of 74.3\% indoors and 46\% outdoors, indicating that the prototype is feasible in the indoor environment. In the outdoor environment, it performs not as expected, as the body electric field will be much weaker than in the indoor environment where a rich amount of powered wires and appliances exiting. Although this preliminary study didn't conclude with a reliable demonstration, it shows the novelty of the body electric field for supplying new sensing approaches, as it still needs to be widely explored.ince the proposed proximity sensing is based on electro-quasistatic coupling, and the human bodies do not only couple with each other and the ground, but also with the environment, factors such as body size, floor material, and nearby electrical devices can influence proximity results. 

Continuing with the idea, the authors explored how this body capacitive sensing-based sonification system could foster basic mechanisms underlying non-verbal social interaction \cite{rizzonelli2022fostering}. Firstly, the authors illustrated that coordination was a crucial primary mechanism for social interaction. Then the sound feedback system was described, including the capacitive sensing platform that takes advantage of the human body’s electric conductivity, and the sound synthesis algorithm that takes the proximity information as the input data. The described was worn on two persons in the form of bracelets and enabled closed-loop auditory interaction between the two persons. Finally, interaction facilitation performance was investigated through behavioral analysis using structured observation, which allows for a quasi-quantitative sequential analysis of interactive behavior. The preliminary inter-observer agreement results showed an increasing percentage of mutual gaze and smile over three sessions of user study, where two participants with a diagnosis of psychosomatic disorders interacted with a trained music therapist. The authors finally argue that the implementation of the sonification system may be fruitful in healthcare contexts and in promoting general well-being.

\subsubsection{Collaborative Physical Games}

Body-to-body interaction can foster not only interaction in therapy but also entertainment, like games. Usage of full-body
interaction in games will benefit the immersion experience of the players. However, this mechanism has not been investigated and implemented in sophisticated gaming apparatus. Based on this observation, Canat et al. \cite{canat2016sensation} designed a device for detecting touch patterns (touch with one-Finger, bro-fist, palm Touch, four-Finger) between players and introduced the game, which is a collaborative game designed to be played with the social touch. The background of the touch sensing is the Swept Frequency Capacitive Sensing technology, as described in some of the previously mentioned works. Unlike single frequency capacitive sensing that only detects whether the touch event occurs due to the change in the received frequency signal, swept frequency capacitive sensing allows to detect touch patterns by observing the shift of the resonant frequency, which is proportional to the system’s total capacitance and changes when different touch patterns are presented. To estimate the contribution of social touch to the overall player experience in collaborative games, the authors conducted a user study with 30 participants and completed a set of questionnaires aimed at measuring immersion levels. As a result, the collected data and observations indicated an increase in general, shared, ludic and affective involvement with significant differences. The result of this work shows that human-to-human touch can be considered a promising control method for collaborative physical games.

\section{Applied Hardware and Data processing pipeline}
\label{sec:pipeline}

% {\bf Frequency the time method can detect static distance, while the charge method only senses the distance in a dynamic state. }

Extracting the context of human activity from body-area capacitive information can be divided essentially into two stages: data acquisition and data processing, depending on how the sensing front end is designed and the complexity of the perceived data. This section presents a summary of applied hardware for data acquisition with the three different sensing sources and the corresponding data processing approach for context extraction presented in published work, aiming to ease other researchers with the same research target by enlightening them for novel sensing and processing ideas on the body-area capacitive signal. 

\subsection{Applied Hardware Sensing Front End}
The human activity caused variation of body-area capacitance provides a promising sensing modality for activity recognition and machine interaction. However, capacitance itself can not be sensed directly by a compact device. Therefore, some other indirect capacitance measurement solutions based on the electrical characteristics of capacitors were proposed to obtain capacitive-related information using specific or general hardware. The applied measurement solutions can be summarized in three types depending on the sensed signal: frequency, current, and time, as those signals are highly relevant to the capacitive variations and easy to read in their specific circuit design.   

\begin{figure}[!t]
\centerline{\includegraphics[width=0.7\columnwidth]{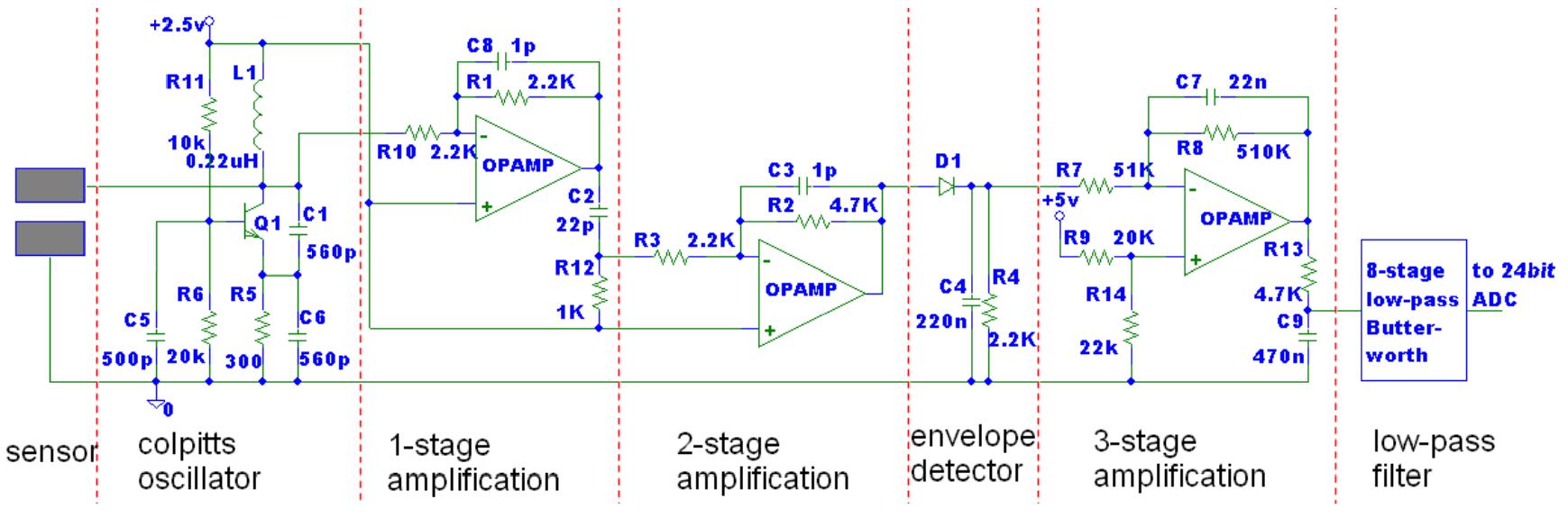}}
\caption{Sensing the body-area capacitance variation by detecting the envelope of an oscillating tank, and example circuit based on the Colpitts Oscillator
\cite{cheng2010active}}
\label{Colpittsoscillator}
\end{figure}

Frequency-based capacitance measurement is the most widely used one. 
As a capacitor is an integral component of many oscillators, there is a definitive relationship between the resonant frequency and capacitance in an oscillator. Therefore, by embedding the human body capacitance into an oscillator, the human body capacitance value can be derived by measuring the oscillator's resonant frequency. Different kinds of oscillator-based sensing front ends were adopted to measure human body capacitance by either reading the envelope of the oscillating AC signal after modulation or counting the frequency of the oscillating AC signal directly.
For example, the Colpitts Oscillator, designed in \cite{cheng2010active} consists of a parallel LC resonator tank circuit with feedback achieved through a capacitive divider. In the case of human body capacitance measurement with the Colpitts Oscillator, the capacitor consists of the electrode, the human body, and the ground. A bipolar junction transistor or operational amplifier is usually used as the oscillator's active stage to generate a sinusoidal waveform signal, as shown in Figure \ref{Colpittsoscillator}. The capacitance variation at the electrode side will result in a variation in both the amplitude and the frequency of the oscillating signal.
In \cite{cheng2010active}, the author adopted the amplitude information for capacitance perceiving. Thus, an envelope detector composed of a diode and low pass filter followed the oscillator. Then, after another signal amplification, the envelope signal was measured by a microcontroller with an ADC interface or an independent ADC module. 
The same workflow is also applied to the swept frequency capacitive sensing technique. By approaching or touching a swept frequency instrumented electrode, a shift of the resonant frequency can be observed, which is proportional to the system’s total capacitance changes. When different approaching or touching patterns are presented, the shift of the resonant frequency will be different and used as signal source for patterns recognition, as described in \cite{harrison2012capacitive, sato2012touche, canat2016sensation}.

\begin{figure}[!t]
\centerline{\includegraphics[width=0.7\columnwidth]{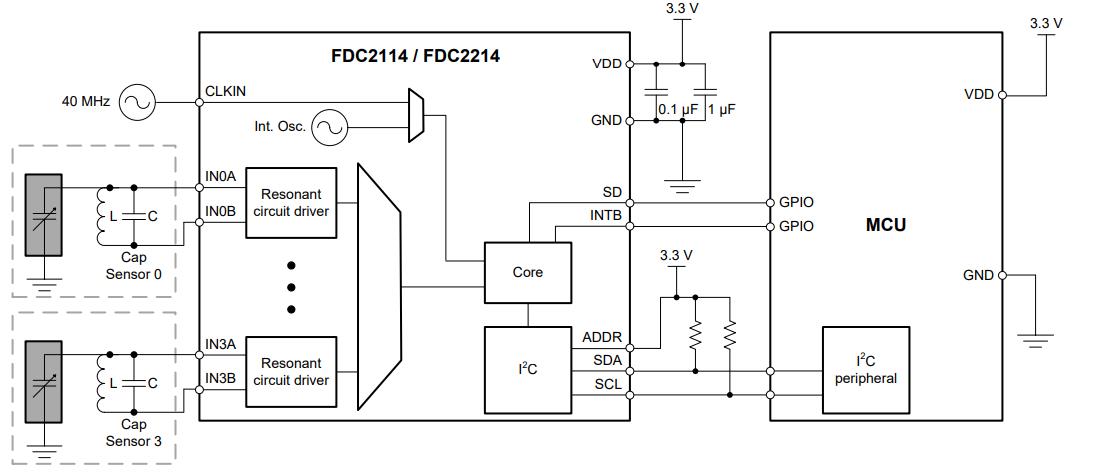}}
\caption{Sensing the body-area capacitance variation by counting the resonant frequency of an oscillating tank, example with FDC2x1x Chip}
\label{FDC2214}
\end{figure}

The other method is to count the frequency number directly, like the capacitance-to-digital converters FDC2x1x from Texas Instrument, as shown in Figure \ref{FDC2214}. Using FDC2x1x for body-area capacitance measurement is based on the fact that a change in capacitance of the L-C tank can be observed as a shift in the resonant frequency. The core of  FDC2x1x reads the frequency, digitizes, and outputs the frequency number, which is proportional to the resonant frequency. An advantage of this chip is that it supports multi-channel capacitance measurement. Besides the FDC2x1x series, the 555 timer chip is also widely used in related works to perceive the body capacitance variation \cite{bian2021systematic, wong2021multi, ramezani2016tagless}. By specific configuration, the output square wave frequency of the 555 timers is determined by the charging and discharging period of the capacitor.

\begin{figure}[!t]
\centerline{\includegraphics[width=0.5\columnwidth]{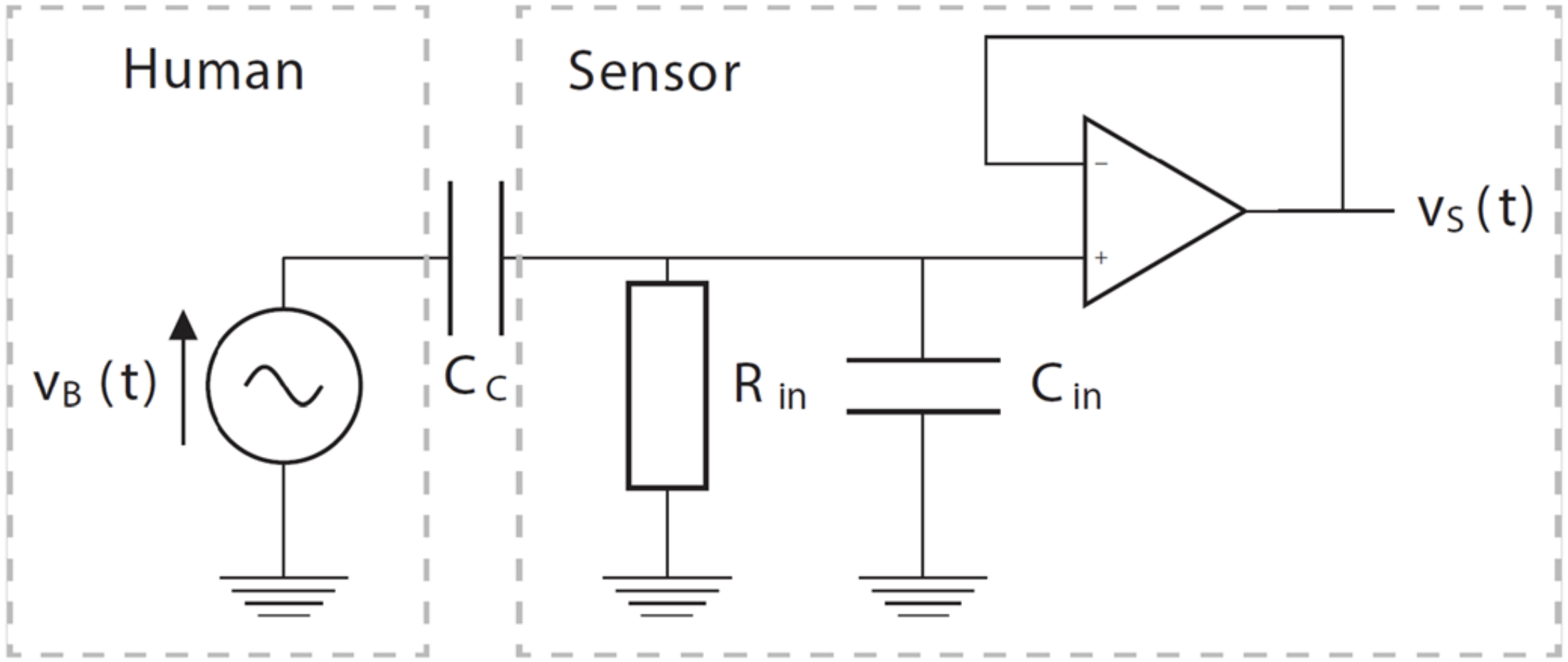}}
\caption{Sensing the body-area capacitance variation with the displacement current, a customized amplification circuit design with high input impedance \cite{fu2019performing}}
\label{current_measurement}
\end{figure}

The second sensing source for body-area capacitance perceiving is the current signal, either in the form of conduction current when the electrode contacts the human skin or in the form of displacement current when the electrode is coupled with the human skin. Figure \ref{current_measurement} shows a straightforward hardware implementation for sensing the body movement caused displacement current with a high input-impedance amplification circuit. 
% C1, C2, and C3 are body-related capacitors, followed by a second-order passive low-pass filter consisting of R1, R2, C4, and C5. A simple voltage divider composed of two high-value resistors is used to convert the current signal to a voltage signal 
$C_{c}$ is a virtual capacitor formed between the human skin and the electrode of the sensing front end. The body movement will result in a bio-potential variation, essentially the flow of charge. This charge flow will be picked up by the electrode when a super high input-impedance ($R_{in}$)  circuit structure is connecting the electrode. Such sensing front end is widely used in related work for sensing the passive body-area electric field \cite{fu2019performing, cohn2012ultra, grosse2016platypus}, benefiting from its low-cost and super low power consumption (in the level of micro-watt). Besides this form, the authors in \cite{bian2019wrist} adopted another current-based sensing structure with only a few tiny-form capacitors and resistors and a high-resolution analog-to-digital converter. The authors explored applications of individual and collaborative activity recognition \cite{bian2022contribution} with the circuit structure and concluded impressive results about the contribution of body-area capacitance in human motion-related recognition tasks. 
%The front end consists of two parts, the human and the sensor circuit, coupled by a capacitor $C_{c}$. The first part is a human body with a varying body voltage $V_{B}$ due to movement. The resistance $R_{in}$ and capacitance $C_{in}$ affect the cutoff frequency of the hardware lowpass filter. The sensed voltage $V_s$ can be measured by the AD module.
% that the AD module can measure; the whole hardware system consumes only several µW power, less than active filter-based sensing
% front ends. 
%This human body capacitance measurement hardware demonstrated remarkable performance in a novel indoor positioning system \cite{fu2019performing}. 
%Besides, there is other applied hardware for current measurement based on the amplifier or specific purpose chip like ADS1298. The core component of this kind of hardware is a transimpedance amplifier converting current signal to voltage signal, which can measure large range current signal by configuring the feedback resistor. 

\begin{table}[!t]
\begin{threeparttable}
\centering
%\begin{threeparttable}
%\renewcommand{\arraystretch}{1.0}
\footnotesize
\caption{Deployed Data Processing Methods From Literature for Single/Multi-Channel One Dimensional Capacitive Data}
\label{tab:DataProcessing}
\begin{tabular}{p{3.0cm} p{4.6cm} p{2.7cm} p{4.0cm}}
\hline

Data Processing Method & Characteristic  & Algorithm & References\tnote{a} \\ \hline
\multirow{5}{*}{Classical machine learning} & \multirow{5}{=}{ Hand-crafted feature extraction to feed more representative information in models; Features in both time and frequency domain; High accuracy with limited volume of data sets}  & Random Forest & T-\cite{matthies2021prototyping}, F-\cite{suzuki2020mouth}, C-\cite{bian2022using}, F-\cite{bian2021systematic}, T-\cite{haescher2015capwalk} \\ 
~ &  & K Nearest Neighbour & C-\cite{cohn2012ultra}, F-\cite{singh2015inviz}, F-\cite{wong2021multi}, F-\cite{wilhelm2020perisense}, F-\cite{ramezani2016tagless} \\ 
~ & & Decision Tree & C-\cite{matthies2017earfieldsensing}, T-\cite{haescher2015capwalk}, F-\cite{cheng2013activity} \\ 
~ & & Support Vector Machine & C-\cite{braun2014towards}, F-\cite{wong2021multi}, F-\cite{ramezani2016tagless}, F-\cite{harrison2012capacitive}, F-\cite{sato2012touche}, F-\cite{cohn2012humantenna}, T-\cite{braun2015capseat} \\ 
%~ &Bayes Net &Time & T-\cite{haescher2015capwalk}  \\

~ & & Naïve Bayes & F-\cite{ramezani2016tagless}, T-\cite{haescher2015capwalk} \\ \hline
\multirow{3}{*}{Deep learning} & \multirow{3}{=}{Better or comparable accuracy with automatic high-level feature extraction; More powerful for large data set with large model}  & Convolutional Neural Networks & F-\cite{bian2021capacitive}, C-\cite{bian2022using} \\ 
~ & & Recurrent Neural Networks & F-\cite{liu2020fpga} \\ \hline
\multirow{3}{*}{Other algorithm} &\multirow{3}{=}{ Features are highly distinguishable from the very limited data; No training is needed; Statistical analysis;}  & Peak detection & F-\cite{cheng2008body}, C-\cite{takiguchi2007human}, C-\cite{bian2019wrist}, F-\cite{bian2021systematic} \\ 
~ & & Maximum likelihood & C-\cite{tang2019indoor} \\
~ & & Weighted-average & T-\cite{braun2011capfloor}\\ \hline

\hline
\end{tabular}

      \begin{tablenotes}
          \item[a] With source signal of T(Time), C(Current) and F(Frequency).
      \end{tablenotes}
      \end{threeparttable}

\end{table}

The time constant used for describing the charging or discharging time on a capacitor is also a valuable parameter for observing capacitance variation. The capacitance value can be derived from the charging and discharging speed under specific conditions. Therefore, the front end based on this principle was also designed to measure the body-area capacitance \cite{zeeman2013capacitive, braun2011capfloor, grosse2013opencapsense}. Such a front end consists of only an RC circuit and a general microcontroller. The microcontroller controls a general-purpose input/output(GPIO) pin to output a new state and then waits for the receive GPIO pin to change to the same state as the send pin, where the RC components are located between the two pins. The capacitance component in the RC circuit is or includes a specific body-area capacitance. Meanwhile, a timer calculates the duration of the time for charging or discharging, which can be used to represent the body-area capacitance variation. The front end, where the time constant is used for capacitance interpreting, is the simplest among the three signal sources.

\subsection{Applied Data Processing Method}
The perceived body-area capacitance data is relatively straightforward, generally in the form of single or multiple channels of one-dimensional continuous data. The dimension depends on the backed hardware configuration, whether using a single electrode as the signal-perceiving unit or using electrode arrays. To check the data processing approaches researchers used to deal with body-area capacitance data, we extracted only the algorithm metric from the previous tables and grouped them into three classes: classical machine learning, neural network-based deep learning, and other straightforward statistical analysis approaches, as shown in Table \ref{tab:DataProcessing}. 
%The variation of human body capacitance is closely related to human activity; therefore, many kinds of human activity can be classified by analyzing and processing the measured body capacitance or capacitance-related value. 
%The proposed data processing methods from existing research papers can be grouped into 

%The data processing pipeline based on classical machine learning and deep learning can be summarized into three steps: feature extraction, classification model training, and activity inference. Compared to inputting raw data directly into the model, there are more benefits to be gained by inputting data features into the model, like reducing data dimension, model size, and increasing inference speed. These data features like variance, max, mean, min, and range are most used to describe raw data. Besides, time serial raw signal can be converted into the frequency domain, and features from the frequency domain like amplitude, kurtosis, skewness and energy can be extracted. Besides, automatically learned features using deep neural networks are also applied in many complex activity recognition tasks.  

The mainly used classical machine learning methods in the literature on body-area capacitance include Random Forest (RF), K Nearest Neighbour (KNN), Decision Tree (DT), and Support Vector Machine (SVM). The one-dimensional raw data is either directly fed into the model for training or, firstly, statistically dealt with for feature extraction. The former is usually used when the critical features are easy to differentiate from the raw data, for example, when the features can be visually captured \cite{wilhelm2020perisense}. In most of the work, feature extraction is performed before feeding data into a classifier. The usage of the feature extraction step increases a feature calculation step but benefits by supplying more representative knowledge to the models for inference. This step is necessary for designing accurate predictive models and, meanwhile, decreasing the time and computation resources needed while making the inference. For body-area capacitive data, the generally extracted features are maximum, mean, minimum, kurtosis, skewness, etc., in the time domain. For data that contains essential context in the frequency domain, the fast Fourier Transform is commonly used to get features in the frequency domain. For example, in \cite{haescher2015capwalk}, five different walking styles(sneaking, normal walking, fast walking, jogging, and walking with weight), which are performed with different speeds and loads, were recognized by trying different classic machine learning models, including Naïve Bayes, Bayes Net, Nearest Neighbor, Decision Tree, and Random Forest. The features extracted are the frequency with the highest amplitude, the highest significant frequency, the spectral centroid, and the signal energy. In \cite{cohn2012ultra}, the authors used six features to classify four activities (rest, arms waving, walking, and jogging) with body capacitance data with a k-nearest-neighbor classifier. Two features are extracted from the power spectral density in the frequency domain, which is often a good indicator of the type of activity. For example, running results in more energy in the higher frequency range than slower activities like walking or resting; and four in the time domain, standard deviation, zero crossings number of the derivative, number of high magnitudes, and magnitude of the first peak of the auto-correlation. An accuracy of nearly 92\% was achieved based on the selected features. Since body-area capacitance data is relatively straightforward, most research reported high classification accuracy with the classical machine learning method. %The classic machine learning model only contains very few hidden layers. 
For example, a random forest model was used to predict 12 facial and head gestures with a reasonably high accuracy of 91.25 \% in \cite{matthies2021prototyping}. In \cite{singh2015inviz}, the authors utilized KNN to classify 16 hand hover and swipe postures with an accuracy of 93 \%. \cite{braun2015capseat} achieved classification precision between 95\% and 100\% for nine static postures and dynamic events using SVM methods. As Table \ref{tab:DataProcessing} shows, SVM is the most selected classifier model by researchers, as it enjoys the advantage of low overfitting risk, especially for small volume data sets, which is the typical case for body-area capacitance-related data sets.

For large-volume data sets, classical machine-learning solutions are often limited by their simple structures. For example, relying on only a few support vectors to determine a multi-classes result, the results are prone to large deviations if there are outliers. A large data set with a deep learning solution will achieve a more generalized result with minimum fluctuations and high-quality interpretations, benefiting from the brain-alike, complex, and intertwined multi-layered neuron network structure. Besides that, deep learning has demonstrated remarkable performance in feature extraction with its shallow layers automatically, while traditional heuristic and hand-crafted feature extraction in the classical machine learning approach heavily relies on professional knowledge and experience. 
Therefore, deep learning methods like Convolutional Neural Networks (CNN) and Recurrent Neural Networks (RNN) are more frequently adopted in body-capacitance-related research. For instance, in \cite{bian2021capacitive}, a CNN-based classifier was applied to recognize seven hand gestures with an accuracy of 96.4 \%. It is worth mentioning that they deployed the CNN model on a microcontroller to realize a real-time inference. Since human activities are made of complex sequences of motor movements, and capturing these temporal dynamics is fundamental for successful HAR \cite{ordonez2016deep}, Recurrent Neural Network is popular for such recognition tasks. Liu et al. \cite{liu2020fpga} proposed a method to detect driver’s movement information using an LSTM neural network, which achieved a leave-one-user-out cross-validation accuracy of 91.3 \%.

Besides the machine learning approaches, statistics- and probability-based algorithms widely recognize activity or interaction in many studies when the capacitive signal is simple and the features are highly distinguishable. For example, in eye-blink detection tasks \cite{luo2020eslucent, liu2022non}, as the variation of sensor signal was very obvious when the user blinked eyes, the authors just used a statistical analysis method to extract the eye blinking information from the pre-processed raw data, like peak detection. The best precision has reached 94 \%. In \cite{bian2021systematic}, the authors also used a peak detection and threshold method for gait partitioning and step counting. The step-counting accuracy was close to perfect. Other algorithms like maximum likelihood \cite{tang2019indoor}, weighted-average \cite{braun2011capfloor}, and independent component analysis \cite{choi2017driver} also demonstrated remarkable performance in human activity recognition tasks. These algorithms often require fewer data to extract context than the former two methods, and they can make a faster inference. However, the performance of these methods is highly dependent on the data complexity and quality.

%"Your noise is my command: sensing gestures using the body as an antenna"
%CapSeat: capacitive proximity sensing for automotive activity recognition
%"CapGlasses: Untethered Capacitive Sensing with Smart Glasses"
%"Capacitive proximity sensing in smart environments"

\section{challenges}
\label{sec:challenges}

Sensing the body-area capacitive signal and signal processing are already mature skills with the development of body-area capacitive sensing exploration in the last decades. However, some issues still block the sensing modality from research to real-life pervasive deployment. Here we briefly discuss the limitation of body-area capacitive sensing, aiming to encourage researchers to explore further to address those challenges with feasible and reliable methods.

\subsection{Robustness}

Lack of robustness is probably the most significant limitation for body-area capacitive sensing for reliable deployment. Several factors result in this limitation. 

The first is the sensor drift, mainly caused by changes in environmental humidity and temperature or the sweat condition on the skin when the electrode is contacted/coupled to the skin. Such conditions will variate the flowing resistance of charge in the form of either direct current or displacement current. Usually, the drift will not be a problem for classification applications, as it changes gradually and can be filtered out. For the other applications that target a regression task, for example, indoor localization with distance information extracted from the proximity reading of a capacitive system, such a misreading will misinterpret as a fake position. 

The second is the omni-sensitivity of the capacitive sensor. In some scenarios, the sensing electrode is supposed to sense activities in a specific direction. For instance, sensing the eye blink or facial expression when attaching the electrode to the frame of a glass. However, since the electrode is omni-sensitive, surrounding proximity from other directions in the vicinity, like approaching of the fingers, will be picked up the electrode and cause confusion in the raw signals. 

The third factor that blocks the robustness of body-area capacitive sensing is the wearing when sensing the body capacitance, which describes the capacitance between the body and mainly the ground. In this case, the shoes play an important role as it is the primary dielectric in between. The bio-potential variation caused by body capacitance change during the body movement will be much weaker when walking barefoot. As verified in \cite{fu2019performing} exploring indoor localization with electric potential sensing, the overall mean positioning error was 18 cm, with a standard deviation of 22.05 cm for recordings without shoes and a mean positioning error of 12.7 cm with a standard deviation of 13.6 cm for recordings with shoes. The different shoes will also cause variated capacitance readings, resulting in an unreliable interpretation. For instance, when authenticating users with body-area capacitance, as stated in \cite{matthies2017capsoles}, the identification on a treadmill while using the same shoe is much less unambiguous than identifying users wearing their own shoe and walking at their individual speed. The same situation also exists in indoor and outdoor environments. Like indoors without shoes, the body capacitance in the outdoor environment also gives much weaker motion-caused capacitance variation, as no noticeable electric wires and appliances exist in the surroundings, and the displacement current faces stronger resistance. This observation can be verified in \cite{bian2022human} when the authors designed the Proxiband and expected social distancing with the whole body passive electric field. The outdoor experiment showed less than half of the intentional social distance break with a true positive judgment, much less than the experiment results in an office building when numerous electric devices are in working status.

\subsection{Generalization}
\label{subsection:Generalization}

Although some research work on body-area capacitive sensing has shown promising results in activity recognition and human-machine interaction, the generalization of the result is still a problem regarding the subject and environment diversity. 

While utilizing the whole body passive electric field as the signal source for a specific application, the body form, like height and weight, is different, or while utilizing a particular body part in a passive or active electric field for gesture or activity recognition, the intrinsic anatomical and physiological factors are also different, which result in different capacitance variation patterns during a movement, as stated in \cite{bian2021systematic, matthies2015botential}. In \cite{grosse2016platypus}, the authors designed a Platypus, a  tag-free system to localize a person and to extract a signature pattern for identification through sensing electric potential changes in human bodies. During the experiments, the authors observed that body electric potential changes are very distinct for different people when taking a step. The reason behind this is, on one side, the external influences, such as footwear and synthetic materials worn; on the other side, the intrinsic physiological, anatomical, and biochemical factors, including the number of active motor units, fiber type composition, blood flow, fiber diameter, depth and location of active fibers and amount of tissue between the surface of the muscle and the electrode. Such intrinsic factors that influence the capacitance-based exploration can be more obviously observed in wrist-based hand gesture recognition, where the subject-dependent accuracy is super high, for example, over 96\% in \cite{bian2021capacitive}, while only 87.24 \% for inter-subject hand gesture recognition reported in \cite{reinschmidt2022realtime} even with more capacitive channels and fused with the accelerometer. This means, for a generalized result with the best performance, person-specific data processing is needed. 

Besides subject diversity, environment diversity also blocks the generalization of body-area capacitance in specific scenarios. The result of some lab-controlled experiment, in some cases, fail when repeating it in a different environment, especially for this body-area capacitive sensing that is highly coupled with environmental configurations. In \cite{cohn2011your}, the authors tried to sense gestures using the body as an antenna to perceive the electric noise from surroundings. In the experiment, the electrical state of the home remained constant throughout the testing session, and the classification works well when the home is in the same state as it was during training; however, large changes in state (i.e., turning on the air conditioning, or all lights in the home) causes the classification accuracy to drop. As there are no simple predictive models to infer what the electrical noise will look like at different locations, it's hard to generalize the novel idea for gesture recognition in different areas. The similar problem is also reported in \cite{braun2015capseat, matthies2021capglasses}. The weak robustness of body-area capacitive sensing to the environment is naturally a shortcoming that blocks the generalization of this novel sensing modality for real-life applications.

%\subsection{Safety}

\section{Outlooks}
\label{sec:Outlooks}

Based on the large scale of targeted applications and the current challenges and development of body-area capacitive sensing in pervasive computing, we proposed several potential directions for future exploration, aiming for a more accurate, robust, easy-of-use sensing with body-area capacitance.

\subsection{Integrated Capacitive Chips}

A reliable body-area capacitive-based sensing application heavily depends on the underlying capacitive front-end design. As summarized in the previous subsection, three kinds of signal sources are commonly used: charging and discharging time, LC- or RC-based resonant frequency, and AC or DC displacement current in contact or coupled form. For the time signal source, any simple GPIOs on a microcontroller could be used to implement the front-end design with simply a few discrete electric components. The capacitance value is represented by recording the charging and discharging time. For the resonant frequency signal source, there have already a few commercially available chips, like the FDC series chips from Texas Instruments released around 2015, which supports multiple channels and multiple precisions, or the traditional 555 timers when configured in the multivibrator mode. The capacitive information is obtained by observing the frequency variation, either by directly counting the frequency number with a timer function on the control unit, or by simply reading through a general peripheral interface from the integrated chip, which already implements on-chip data post-processing with a capacitance-to-digital converter. The simplicity and flexibility of the resonant frequency-based capacitive sensor allow it to be easily incorporated into various systems that require proximity sensing. For the displacement current signal source, capacitive monitoring is accomplished by observing the charge flow in the form of potential variation on an electrically connected high-input impedance circuit. An analog-to-digital converter is needed for numerical reading. Compared with the time and frequency signal source, which can continuously interpret a capacitance value, no matter whether the sensed object is in a static or dynamic state, the current monitoring solution only senses the capacitance information when there is a variation in the object state, like moving the body part. Since only the variation of the object or environment will cause the charge flow on the electrode. This character makes the current-based solution super sensitive, which can be an advantage since it enables the front end to sense full-body movement regardless of the sensor placement position, which means, for example, a wrist-worn sensor can sense the leg movement when the wrist is in a static state, which is attractive as no other wearable sensors having the same sensing ability, including the pervasively embedded inertial measurement unit. On the other hand, this can be a disadvantage since any environmental variation, like the intrusion of a second body, will result in unwanted signals when targeting the movement of the sensor-worn subject. Another impressive advantage of the current-based solution is the super low power consumption at the level of uW compared with the other two at the mW level.

%\begin{figure}[!t]
%\centerline{\includegraphics[width=0.4\columnwidth]{Figures/Plessey.png}}
%\caption{Electric Potential Integrated Circuit (EPIC) sensors from Plessey. Source: Plessey Semiconductors Ltd}
%\label{Plessy}
%\end{figure}

%\begin{figure}[!t]
%\centerline{\includegraphics[width=0.4\columnwidth]{Figures/Qvar.png}}
%\caption{LIS2DUXS12: Ultralow-power accelerometer with Qvar. Source: ST Microelectronics}
%\label{Shielding}
%\end{figure}

\begin{figure}
%\centering
\begin{minipage}{.45\columnwidth}
    \centering
    \includegraphics[width=0.5\columnwidth, height = 3.5cm]{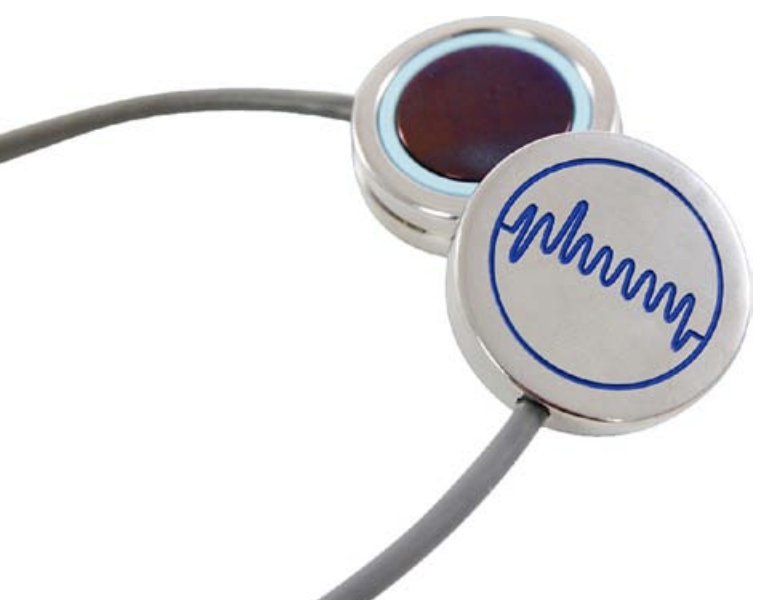}
    \caption{Electric Potential Integrated Circuit (EPIC) sensors from Plessey. Source: Plessey Semiconductors Ltd}
    \label{Plessy}
\end{minipage}%
\quad
\begin{minipage}{.45\columnwidth}
    \centering
    \includegraphics[width=0.7\columnwidth, height = 3.5cm]{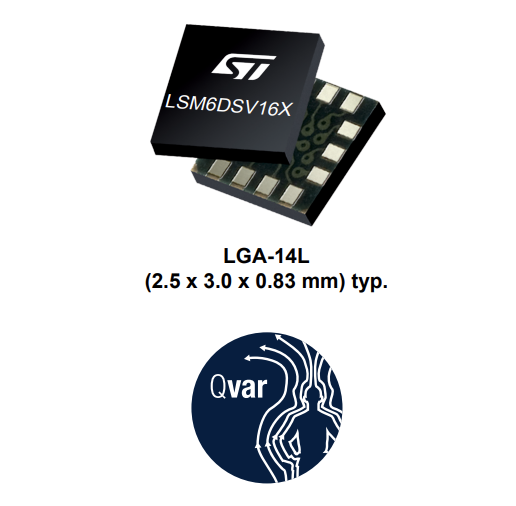}
    \caption{LIS2DUXS12: Ultralow-power accelerometer with Qvar. Source: ST Microelectronics}
    \label{Qvar}
\end{minipage}
\end{figure}

Despite of impressive properties of the current-based solutions, most of the research work designed their own sensing front-end composed of discrete components for body-area displacement current monitoring with target application of motion sensing \cite{bian2022using}, physiological signal recording \cite{tang2022high}, facial expression recognition \cite{matthies2017earfieldsensing}, indoor localization \cite{fu2019performing}, etc. The most crucial factor for a successful design is the super high input impedance, in order to pick up the tiny charge flow signal on the deployed electrode. In 2011, Plessey Semiconductors Ltd announced the availability of commercial samples of its electric potential integrated circuit (EPIC) sensors \cite{bogue2012plessey}, as Figure \ref{Plessy} shows, which arose from a collaboration between Plessey and the University of Sussex, UK. The EPIC sensor is a wideband (quasi DC to 200 MHz) ultra-high impedance sensor capable of detecting spatial potential and electric field by collecting the charge status on the electrodes. The release of the sensor targeted medical, home health, security, non-destructive testing, geophysical surveying, and human-machine interfacing applications. However, this product was not mass-produced and commercially unavailable. In 2022, ST Microelectronics released two MEMS sensors (the inertial sensor LSM6DSV16X as Figure. \ref{Qvar} shows and the barometer ILPS22EHQV) embedded with Qvar module, which stands for Electric Charge (Q) Variation (var) and composed of basically an instrumentation amplifier with ultra-high input impedance,  enabled sensors to detect the differential electric potential variation induced on the electrodes. The Qvar module could work in two modes, contact or coupled with human skin for enhanced human activity detection and deployed in the environment for presence sensing. Some works have already been explored based on the Qvar module, like hand gesture recognition \cite{reinschmidt2022realtime}, and non-contact ECG perceiving \cite{dheman2022cardiac}. Considering its pervasive usage scenarios and advantages in power consumption and fusion design with other sensing modalities, more Qvar-based applications will be developed in the future. 

In the hardware aspect of body-area capacitive sensing, designing an integrated capacitive sensor featuring ultra-low power and adaptive sensitivity while maintaining a robust sensing ability is a valuable future work, and it will enable a new era of wearable motion sensing. The current-sensing-based capacitive sensors, like Qvar, are sensitive to surrounding noises despite their ultra-low operating power. The Time- and Frequency-based capacitive sensors are more resistive to unexpected intrusion but still consume power at the level of several $mW$, which is already power-saving but still makes them non-comparable to the state-of-art inertial measurement sub-$mW$ sensors for wearables. Considering the fact that inertial sensors are the only motion-sensing component and face certain limitations (mostly only sense the motion pattern of the deployment site, like the wrist, and will lose information for other body sensing when worn on wrist), body-area robust and low-power capacitive sensors will be a good complementary solution.

\subsection{Active Shielding}

\begin{figure}[!t]
\centerline{\includegraphics[width=0.5\columnwidth]{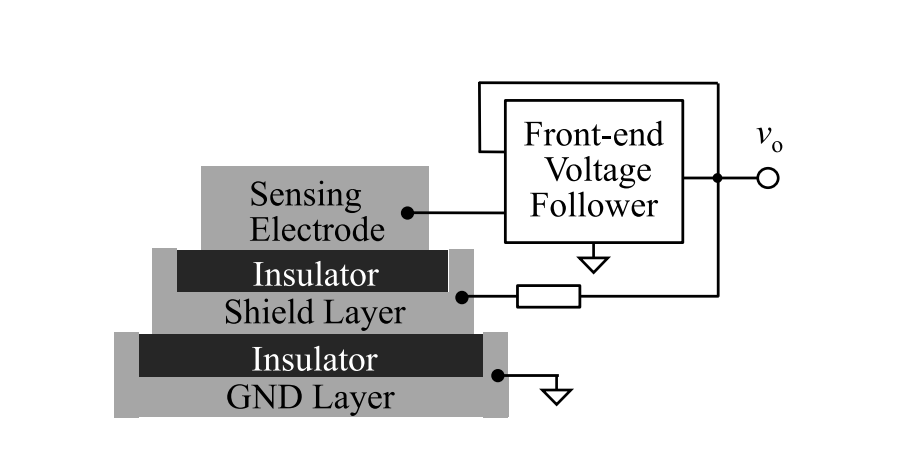}}
\caption{Cross-sectional diagram of the electrode and connection to a
front-end voltage follower \cite{takano2017non}. The sensing electrode is doubly shielded by the third and fifth layers. The output of the voltage
follower is fed back to the third layer. The fifth layer serves as
a ground plane.}
\label{Shielding}
\end{figure}

\begin{figure}[!t]
\centerline{\includegraphics[width=0.5\columnwidth]{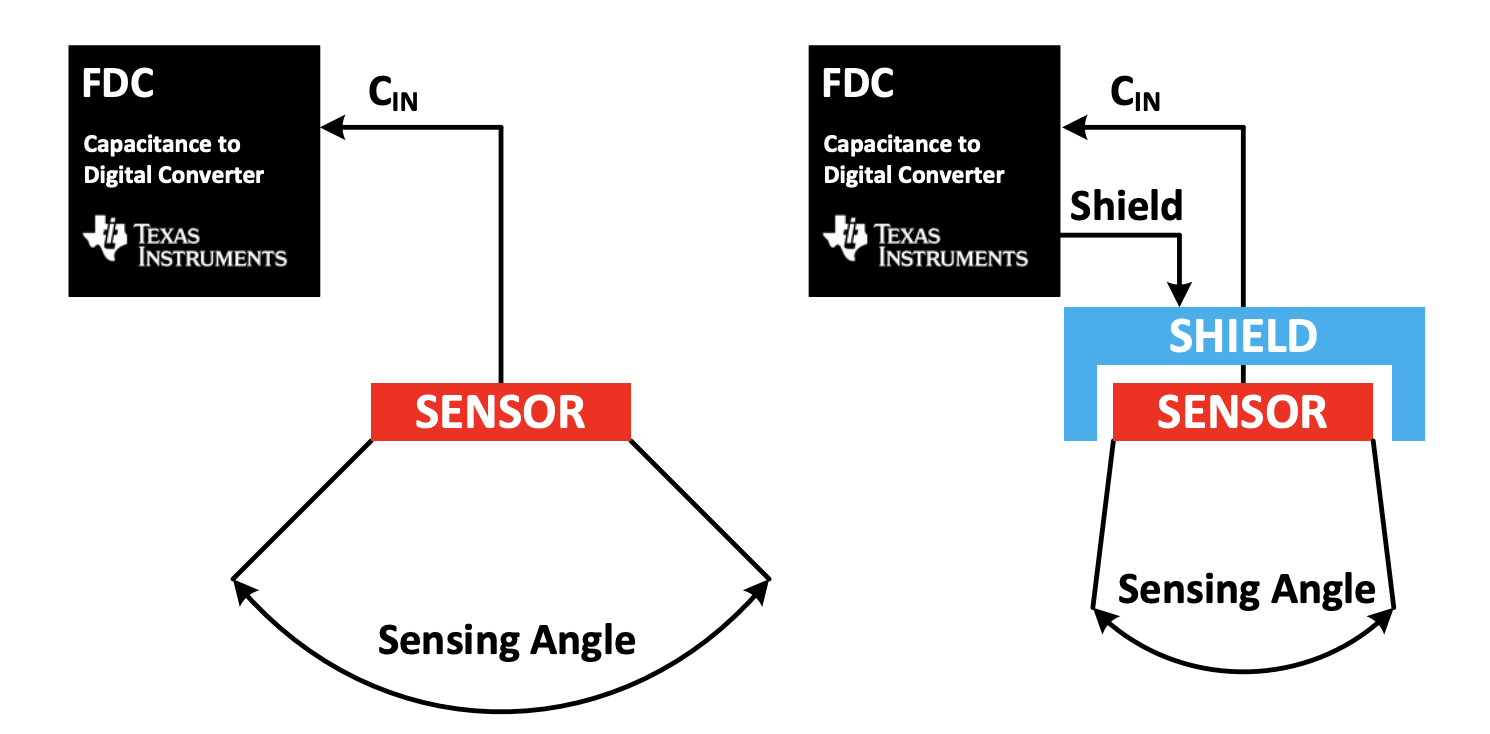}}
\caption{Sensing angle is decreased by the use of a shield \cite{wang2015capacitive} (FDC chips from Texas Instruments).}
\label{Shielding_TI}
\end{figure}

Although body-area capacitive sensing is advantageous in low power consumption and pervasive usage scenarios in wearable deployment, certain aspects of using this technology require specific attention, like environmental interference and parasitic capacitance. Active shielding is an efficient approach to help mitigate the environmental interference that causes artifact readings. A shield paired with the sensing electrode can provide a barrier against external interference so that a better signal-noise ratio is achieved, and only the interesting area will be sensed. Some of the capacitive sensing works have already implemented active shielding in their front-end design. 

In \cite{takano2017non}, the authors designed a double-shield electrode consisting of five layers (as Figure \ref{Shielding} depicts) to improve tolerance against disturbance for non-contact simultaneous measurements of electrocardiogram and respiratory movements with an instrumented sheet. Where the first layer is a sensing layer, the fifth layer is a ground plane, and the output from the front-end voltage follower is fed back to the third layer to configure a driven shield, which can, firstly, suppress current leakage from the sensing electrode to the fifth layer to improve the signal-to-noise ratio; Secondly, keep the sensing electrode being in high impedance in the same manner of guard ring to resist the disturbance. The functionality of the doubly shielded electrode with the driven shield and the ground layers was demonstrated in the study of an in-vehicle ECG device, where the tolerance against vibratory and electromagnetic disturbances is improved compared to the single-layer case.

\begin{table*}[!t]
%\captionsetup{justification=centering,labelsep=newline,textfont={footnotesize,sc},labelfont=footnotesize}
\centering
\begin{threeparttable}
\caption{Comparison of Wearable IMU and Body Capacitance Sensing}
\label{Comparision}
\begin{tabular}{ p{1.2cm} p{0.5cm} p{0.8cm} p{1.5cm} p{0.9 cm} p{2.9cm} p{1.6cm} p{1.4cm} p{1.2cm}}
\toprule
Modality & size & cost & power cost\tnote{a} (1.8V Input)  & output & circuit form & attach position on body & surround- ing sensitive & Robust- ness\\ 
\midrule
IMU & small & Euros & 0.65 mA & digital & integrated circuit chip & motion part & no & strong\\
\midrule
Body Capacitance & small & tens of cents & 15 uA & analog / digital & discrete components / newly integrated circuit chip & motion/static part & yes & weak\\ 
\bottomrule
\end{tabular}
      \begin{tablenotes}
          \item[a] Take LSM6DSV16X as example.
      \end{tablenotes}
      \end{threeparttable}
\end{table*}

Some of the commercially available capacitive chips already implemented shield skills in their system design, like FDC1004 from Texas Instruments, where an extra pin on the small footprint is designed to protect the capacitance input pin from unexpected electromagnetic interference and help focus the sensing field. Figure \ref{Shielding_TI} depicts the variation of the sensing angle with/without a shield near the sensing electrode with the finite element analysis simulations. The sensing angle without a shield picks up any stray interference within the field-line vicinity. The sensing angle with a shield depends on how large the shield is compared to the sensor and how close the shield is to the sensor. The experimental results quantify a trend between shield size, sensitivity, and interference, even though the relationship is non-linear.

Active shielding in capacitive sensing has been explored widely in the last decade, mostly in conjunction with specific applications \cite{hwu2013shielding, reverter2006stability}. As the body-area capacitive character is ubiquitous and complex when interacting with the instrumented or non-instrumented environment, a focused sensing area, higher signal-to-noise ratio, and accurate capacitive reading are required for a robust and reliable sensing design. However, from the extensive survey of published related works, we found that active shielding is still seldom adopted in customized front-end design. Thus the related works are facing the problem of robustness and sometimes generalization, especially caused by unexpected environmental disturbance, like body capacitive-based social distancing \cite{bian2022human}, capacitive glasses for facial gesture recognition \cite{matthies2021prototyping}, etc. Nevertheless, considering the positive support of active shielding in capacitive sensing and almost zero cost in power consumption, size, and price, we believe that active shielding will be considered more in the future body-area capacitive sensing design. 

With a directional sensing feature supported by active shielding on a body-area capacitive sensing system, plenty of existing body-area capacitive works that are facing the trouble of non-robust can be re-explored with a much more feasible solution. A straightforward example is the body-capacitance-based step counting, which could potentially supply the near-truth step number when the inertial sensor on the wrist loses the step counting functionally, for example, when pushing a shopping cart in a supermarket. However, the current wearable solution with body capacitance senses not only the steps of the subject but also the steps of passersby. A directional on-body capacitive sensor will limit the sensing electrode only response to the electric field variation of the subject, thus compressing the external, noisy electric field. Besides the wearable applications, directional capacitive sensing will also bring benefits to capacitive-based biological signal monitoring, like the EEG, in the hospital environment, where lots of powered instruments are installed and act as noise sources for capacitive-based sensing systems.

\subsection{Sensor Fusion}

Considering that current body-area capacitive sensing is not robust against unexpected interference from surrounding conductive intruders, especially when using the displacement current as the signal source to monitor the capacitance variation, the capacitive sensing modality alone is incapable of extracting high-level context from the raw reading. A lot of previous work has already demonstrated this shortcoming \cite{liu2022non, dheman2022cardiac}. In \cite{liu2022non}, the authors deployed the electrode on a general glass frame, and the FDC2214 capacitive to digital converter chip was adopted to sense the eye blinking, with the background that the regular action of the upper eyelid will result in regular capacitive value on the electrode. Although the real-life experiment shows impressive detection accuracy in both intentional and involuntary eye blink, the authors reported that intensive body actions like running cause the glass vibration relative to the head, which results in a more significant signal than the blink-caused signal. To recognize the blink component and vibration component in the raw reading, an inertial sensor fusion is a promising solution. Some other works explored sensing fusion to find the contribution of capacitive sensing \cite{reinschmidt2022realtime, bian2022contribution, bian2022using}. In \cite{reinschmidt2022realtime}, the charge sensing module was demonstrated to be able to improve the accuracy of IMU-alone hand gesture recognition by more than 10\%; In \cite{bian2022contribution}, the body capacitance variation information supplied better results over the IMU for workouts counting(0.800 vs. 0.756 when wearing the sensors on the wrist), and improved the recognition of collaborative activities with an F-score over a single wrist accelerometer approach by 16\%. As can be seen, fusing capacitive sensing and inertial measurement sensing is meaningful and promising, especially in motion pattern recognition scenarios. 

Table \ref{Comparision} listed the crucial characteristics of the IMU sensor and body-area capacitive sensor when targeting wearable motion sensing. Both of them have a small form factor to fit into wearable devices. At the same time, the body capacitive sensor outperforms in cost (nothing more than a few discrete components and an amplifier), power consumption (with reference from LSM6DSV16X), attach position (full-body motion sensing ability for body capacitive sensor, while the IMU has to be deployed on the moving body part), and sensing field (being able to sense the surrounding variation). However, the body capacitive sensing is weak in robustness, which is crucial for a reliable application. Regarding the circuit form, the IMU sensor has long been in the form of a tiny MEMS structure, while capacitive sensors are primarily in discrete form when using the displacement current as the signal source. The only commercially available integrated charged-based capacitive chip is the Qvar-integrated MEMS sensors (fused with inertial sensor and barometer sensor) from ST Microelectronics released in 2022, as described before. This will bring impressive results in exploring body capacitive-based motion and presence sensing in a sensor fusion approach for robust results.

Besides the fusion with inertial sensors, the body-area capacitive sensing could also be used as a triggering signal for other sensors like camera, in some edge, battery-powered embedded systems that require long-term environmental monitoring, taking advantage of ultra-low power consumption and meter-level intrusion sensing range characters of body-area capacitive sensors. With proper configuration, an instrumented capacitive sensor can sense the proximity of a human body in meters away with a power level $uW$.

\subsection{Subject-dependent Continuous Learning}
As mentioned in \ref{subsection:Generalization}, the generation of human activity recognition-based capacitive sensing is still a tough problem because of the subject and environment diversity. Most of the traditional offline machine learning methods are often designed to learn a model from the entire training data set at once \cite{hoi2021online}, which means that there will not be new knowledge that can be obtained by this trained model in the future.
Therefore, the recognition accuracy could decay over time due to the variation in environment and users' behavior in real-life application scenarios. 
To keep a reliable activity accuracy for all users over a long time, subject-dependent computing is an important work direction for practical real-world applicability of capacitive sensing-based technology. Online learning and incremental learning provide a promising solution for subject-dependent computing, as the model based on online learning has the ability to update the model from new data continually regarding different subjects and environments. Online learning solution consists of two main components: model inference and model training. 
The model training process is often performed if the recognition accuracy degradation is detected or new activity is added.
For example, in \cite{yu2021fedhar} the authors proposed a personalized federated HAR framework and developed an unsupervised gradient aggregation strategy to overcome the concept drift and convergence instability issues in online federated learning process. The proposed framework demonstrated an additional 10\% improvement across all metrics on average. 
In \cite{disabato2022tiny}, the authors introduced a tiny machine learning model for concept drift (TML-CD) solution based on deep learning feature extractors and a k-nearest neighbors (k-NNs) classifier. Three adaptation mechanisms, such as passive update, active update, and hybrid update were employed to correct the KNN classifier. In active update mechanisms, the adaptations were carried out at each new incoming supervised sample without detection of an accuracy change. There is a change detection test in active mechanisms, the update process was executed when a change is detected. The hybrid mechanism is a combination of the first two. Noteworthy, the proposed solution can be deployed on a resource-constrained microcontroller. Different from online learning based on the classical machine learning method, the training process of the model based on deep learning techniques requires much more computation and memory resources than inference, which is a considerable challenge when deploying the online learning model on edge devices(on-device learning). Usually, the training process is completed on the cloud \cite{awan2015subject,chen2017robust}. However, the data exchange can lead to high power consumption and privacy issues. Training on edge devices is becoming a promising method for online learning to overcome these above problems. To overcome the resource-constrained problem of on-device learning, many solutions were proposed, like reducing the number of trainable parameters \cite{han2015deep, cai2019once} and reducing the activations \cite{cai2020tinytl}. Recently, an algorithm-system co-design framework to make on-device training possible with only 256KB of memory was proposed in \cite{lin2022device}. The framework enables IoT devices not only to perform inference but also to continuously adapt to new data for on-device lifelong learning, which will be an important cornerstone for subject-dependent continuous learning on edge devices.

In the case of body-area electric field sensing, most of the data in the existing plentiful explorations are processed with supervised classical machine learning or deep neural networks methods, which means the trained model parameters are fixed without considering future data shift caused by environmental variation or unregistered users. As body-area electric field sensor data is highly personalized, online learning with an incremental training mechanism could make use of the incrementally collected unlabeled user samples to actively adjust the model parameters. Such an adaptive online learning method will benefit most of the model-based exploration in the field of HCI and HAR.

\section{Conclusion}
\label{sec:Conclusion}

Body-area capacitive/electric field sensing is a promising sensing modality for human activity recognition and human-computer interaction backed by the property that the body is an ideal conductor. Due to this feature, the body can, first, radiate a passive electric field from the body to the surroundings, and second, deform an existing electric field in an instrumented environment. Previous works based on this property have shown a significant increasing trend in the last decade, including motion sensing, physiological sensing, assisted living, entertainment, etc. Some designs have reported 100\% activity recognition accuracy (for example, in the seat posture detection application), and a uW-level of power consumption for wearable motion sensing (for example, in the task of workouts recognition), which is over ten times power saving than the traditional wearable motion sensor IMU. However, despite the pervasive exploration of the body-area electric field, a comprehensive summary doesn’t so far exist for an enlightening guideline. This paper fills in the gap by comprehensively summarizing the existing related works so that researchers can have a systematic overview of the current exploration status of body-area capacitive sensing. For this purpose, we categorized the topic into three classes:  body-part capacitive sensing, whole-body capacitive sensing, and body-to-body capacitive sensing, and enumerated broadly the published works within each category. An in-depth description of the underlying technical tricks is given, including hardware implementations, applied algorithms, targeted applications, performance, and limitations. Especially, we summarized the sensing source signals for body-area capacitive applications, namely the time, frequency, and current. We also discussed their suited usage scenarios, aiming to provide essential information so that other researchers can decide if and in what capacitive sensing form is suitable for their specific applications. Finally, we analyzed the challenges for the massive deployment of the body-area capacitive sensing technique in practical applications like the weakness in robustness, and proposed several directions for a more accurate, robust, easy-of-use sensing with body-area capacitance, aiming to encourage researchers for further novel investigations considering the pervasive and promising usage scenarios backed by body-area capacitive sensing.

%\begin{acks}
%To Robert, for the bagels and explaining CMYK and color spaces.
%\end{acks}

\bibliographystyle{ACM-Reference-Format}
\bibliography{sample-base}

\appendix
 \section{History}
 \label{History_appendix}

\begin{figure}[!t]
\centerline{\includegraphics[angle=0, width=1.0\columnwidth, height=18cm]{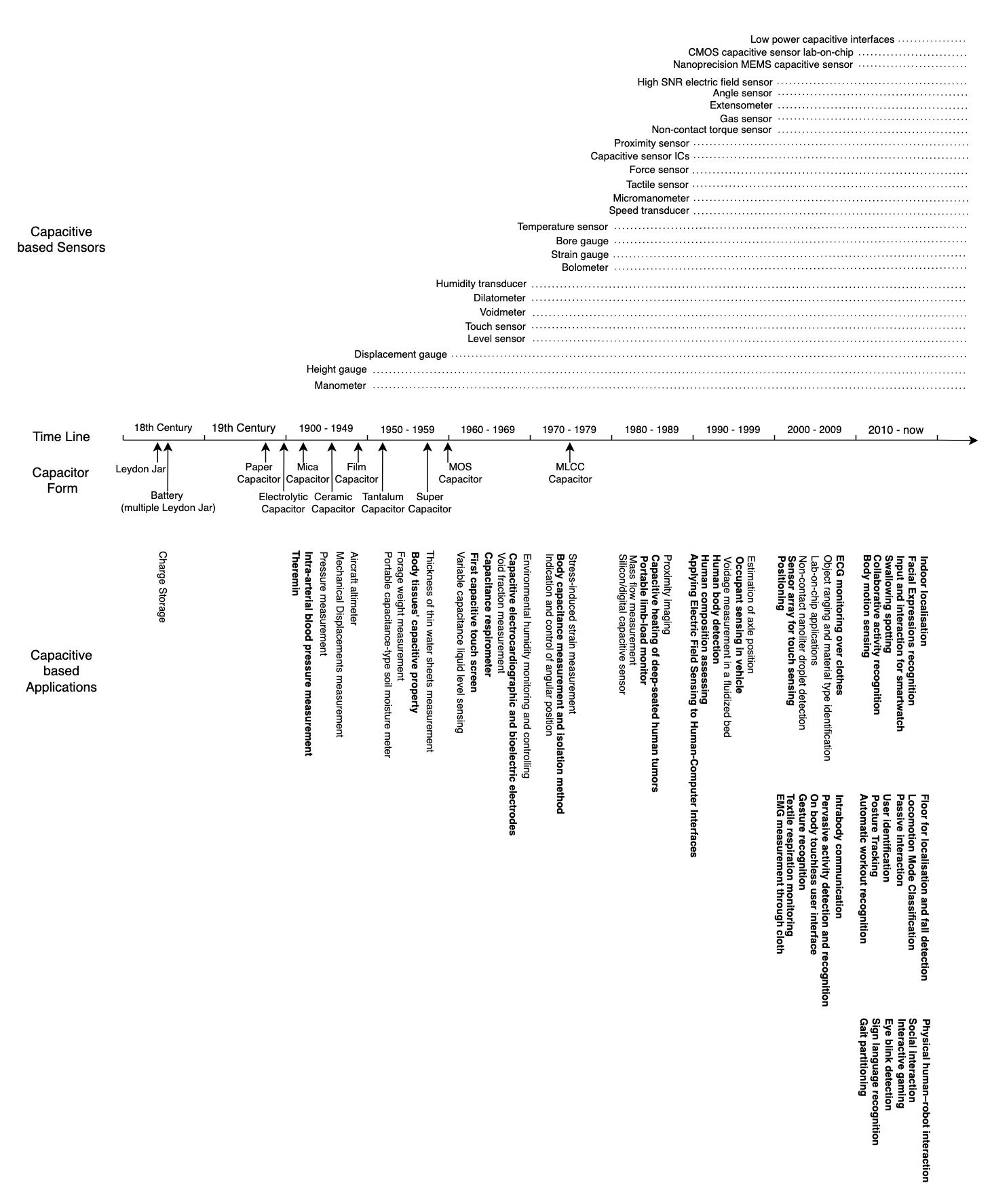}}
\caption{History of the capacitor forms, capacitive based sensors and the development of capacitive based application (body-area capacitive sensing in bold form)}
\label{History}
\end{figure}

Capacitive sensor was born in the pioneering days of electricity and was used to store the charge in the form of Leyden Jar \mbox{\cite{heilbron1966gm}}, as electricity was thought to be fluid in the 18th century. In the 1830s, Michael Faraday did experiments that determined that the material between the capacitor’s plates had an effect on the quantity of charge on the capacitor’s plates. For this work, the unit for capacitance is called the farad. In the late 19th century, paper capacitors and electrolytic capacitors were invented and were used from the early 20th century, for example, as decoupling capacitors in telephony. During the second and third industrial revolutionaries, different forms of capacitors were invented to satisfy the rapid development of industry, like the ceramic capacitor used in the resonant circuit and the tantalum capacitor used as a reliable low-voltage support capacitor to complement the newly invented transistor. Meanwhile, capacitive-based sensors were explored to measure the height (height gauge), pressure (manometer), displacement (displacement gauge), etc. 
From the early 20th century, plentiful capacitive-based sensors were developed and optimized to sense the physical world, especially driven by the silicon industry and the MEMS technique \mbox{\cite{bogue2007mems}}. 
With the development of capacitive sensors, a rich amount of capacitive-based applications were explored, like the measurement of the thickness of thin water sheets \mbox{\cite{black1959capacitance}}, liquid level \mbox{\cite{wilner1960variable}}, humidity \mbox{\cite{misevich1969capacitive}}, displacement \mbox{\cite{todd1954capacitance}}, etc. Meanwhile, the body-area capacitive sensing also started to be explored, like the capacitive blood pressure manometer \mbox{\cite{tompkins1949new}}, which senses the capacitance variation caused by the varied blood pressure in the arteries.
The first body capacitance-enabled consumer device is the special music instrument Theremin \mbox{\cite{nikitin2012leon}}, invented by the 23-year-old Leon Theremin in 1919 by accident. The story of Theremin, as a key milestone, has been widely described in body-area electric field-related works in the background and historical sections \mbox{\cite{smith1996field, grosse2017finding, braun2015capacitive}}. "He was working in a laboratory in Russia as a young scientist, and he was actually working on a gas meter to measure the density of gases, so as he brought his hand closer to the gas meter, he heard kind of a higher squeal. And as he brought his hand back to his body and away from the machine, it was a slower squeal." wrote Glinsky in \mbox{\cite{glinsky2000theremin}}. 
Just as in the pioneering days of aviation, when humans made their own planes out of wood and canvas and struggled to leap into the air, not understanding enough about aerodynamics to know how to stay there, Theremin experienced the same period. 
With decades of development, electrical Theremin was invented \mbox{\cite{kuik2004digital}}, and the behind principle has been evidently verified \mbox{\cite{skeldon1998physics}}: the human body capacitance (or the static body electric field), namely the capacitance between the human body and the surroundings. 
The two metal antennas on the instrument sense the relative position of the thereminist's hands and control oscillators for frequency with one hand and amplitude with the other. The electric signals from the theremin are amplified and sent to a loudspeaker. 
Similar body area capacitance or body-area electric field could also be found in special fishes like the gymnarchus niloticus \mbox{\cite{lissmann1958mechanism, bullock1982electroreception}} with a more sensitive way: the potential distribution over the surface of the fish is detected by a series of receptors; this information is then interpreted to indicate the position of objects with a conductivity differing from that of water. 
In 1965, E.A. Johnson invented what is generally considered the first finger-driven capacitive touchscreen \mbox{\cite{johnson1965touch}}, utilizing the conductance/capacitance property of the finger. Besides that, researchers and engineers also explored body capacitance-based respirometer \mbox{\cite{barrow1969capacitance}},  the limb loads exerted by patients \mbox{\cite{miyazaki1986portable}}, etc.. The body-area capacitance measurement \mbox{\cite{jonassen1998human}} and isolation method \mbox{\cite{forster1974measurement}} for patient protection were also investigated, as in the patient environments, the presence of electrostatic fields arising from unshielded wiring will result in an isolated patient exhibiting capacitance between the earth and the active leads. 

Starting from the 21st century, body-area capacitive sensing showed an explosive development benefitting from its pervasiveness and the emerging novel and high-precision sensing approaches, such as the resonator-based capacitance chips like FDC2x1x from Texas Instruments(TI), and the charger variation-based capacitance chips like QVAR from STMicroelectronics. Since capacitive sensing is a low-power, low-cost, contactless sensing technique, it was applied to applications like biophysical signal monitoring \mbox{\cite{yama2007development}}, position tracking \mbox{\cite{osoinach2007proximity}}, activity classification \mbox{\cite{cheng2010active}}, and intrabody communication \mbox{\cite{shinagawa2004near}}, etc. 
As a summary, Fig. \ref{History} depicts the history of capacitance development, including the different capacitor forms, capacitive-based sensors, and their backed applications, with body-area applications in bold format.

\end{document}